\documentclass[AMA,LATO1COL]{WileyNJD-v2}
\usepackage[utf8]{inputenc}
\usepackage{amsmath,amssymb,amsfonts}
\usepackage{graphicx}
\usepackage{textcomp}
\usepackage[export]{adjustbox}
\usepackage{todonotes}
\usepackage{hyperref}
\usepackage[labelfont=bf]{caption}
\usepackage{subcaption}
\usepackage{paralist} 
\usepackage{tabto}
\usepackage{float}
\usepackage{flushend}

\usepackage{csquotes}

\usepackage{multirow}
\usepackage{booktabs}
\usepackage{colortbl}
\usepackage{dirtytalk}

\usepackage{tcolorbox}

\newcommand\says[1]{\emph{``#1''}}

\newcommand\papertitle{Design Thinking and Creativity of Co-located vs. Globally Distributed Software Developers}

\articletype{Article Type}

\received{26 April 2016}
\revised{6 June 2016}
\accepted{6 June 2016}

\begin{document}

\title{\papertitle}

\author[1]{Rodi Jolak}

\author[2]{Andreas Wortmann}

\author[3]{Grischa Liebel}

\author[4]{Eric Umuhoza}

\author[5]{Michel R.V. Chaudron}

\authormark{RODI JOLAK \textsc{et al}}

\address[1]{\orgdiv{Department of Computer Science and Engineering}, \orgname{Chalmers \textbar \space Gothenburg University}, \orgaddress{\country{Sweden}}}

\address[2]{\orgname{RWTH Aachen University}, \orgaddress{\country{Germany}}}

\address[3]{\orgdiv{School of Technology}, \orgname{Reykjavik University}, \orgaddress{\country{Iceland}}}

\address[4]{\orgname{Carnegie Mellon University-Africa}, \orgaddress{\country{Rwanda}}}

\address[5]{\orgname{Eindhoven University of Technology}, \orgaddress{\country{Netherlands}}}

\corres{*R. Jolak, \email{rodi.jolak@cse.gu.se}}

\presentaddress{Department of Computer Science, Chalmers \textbar \space Gothenburg University, 41296 Gothenburg, Sweden}

\abstract[Summary]{
\emph{Context}: Designing software is an activity in which software developers think and make design decisions that shape the structure and behavior of software products. Designing software is one of the least understood software engineering activities. In a collaborative design setting, various types of distances can lead to challenges and effects that potentially affect how software is designed.

\noindent \emph{Objective}: To contribute to a better understanding of collaborative software design, we investigate how geographic distance affects its design thinking and the creativity of its discussions.

\noindent \emph{Method}: To this end, we conducted a multiple-case study exploring the design thinking and creativity of co-located and distributed software developers in a collaborative design setting.

\noindent \emph{Results}: Compared to co-located developers, distributed developers spend less time on exploring the problem space, which could be related to different socio-technical challenges, such as lack of awareness and common understanding. Distributed development does not seem to affect the creativity of their activities.

\noindent \emph{Conclusion}: Developers engaging in collaborative design need to be aware that problem space exploration is reduced in a distributed setting. Unless distributed teams take compensatory measures, this could adversely affect the development. Regarding the effect distance has on creativity, our results are inconclusive and further studies are needed.

}

\jnlcitation{\cname{
\author{R. Jolak}, 
\author{A. Wortmann}, 
\author{G. Liebel}, 
\author{E. Umuhoza}, and 
\author{M. R. V. Chaudron}} (\cyear{2020}), 
\ctitle{\papertitle}, \cjournal{J. Software: Evolution and Process}, \cvol{2020;00:1--6}.}

\maketitle

\footnotetext{\textbf{Abbreviations:} SE, Software Engineering; UI, User Interface}

\begin{tcolorbox}
\textbf{Self-archiving note:}\\
This is a pre-peer-review version of an article published in Wiley Journal of Software: Evolution and Process. The final version is available via \url{https://dx.doi.org/10.1002/smr.2377}
\end{tcolorbox}

\section{Introduction}
\label{sec:intro}

When designing software, developers, together with other stakeholders, explore the interplay of problem and solution space.
That is, they creatively ponder-, make- and refine decisions that ultimately shape the final structure and behavior of the software product \cite{jolak2017dissecting}.

Throughout the software engineering (SE) life-cycle developers with various roles \cite{petre2016software} jointly are \emph{designing} the system.
For example, developers do not only elicit requirements, they actually design the requirements by discussing and shaping these requirements with the contributing stakeholders. 
Developers also design their code by modularizing-, composing-, analyzing and evaluating source code. 
Similarly, developers design use cases, user interfaces, APIs and test cases. 
All of these activities demand creativity. 

To handle
complex problems, expert developers intuitively practice \emph{design thinking} \cite{cross2011design}, which is a cognitive style, a mindset that helps developers in problem solving.
In \emph{design thinking}, developers explore the problem and solution spaces separately, and iteratively align the two.
This process happens even if developers are not specifically trained in \emph{design thinking}.
While thinking about the design, developers might have sudden events of insight, leading to the establishment of key design concepts or decisions. 
These events are considered as indicators of creativity in the design process~\cite{dorst01}.

Globally-distributed projects are becoming the norm in SE \cite{ebert2016global} and they raise social-, technical-, and organizational challenges \cite{damian2006guest}.
Consequently, it is quite likely that software design activities are affected by these as well.
In particular, it is unclear to what extent the iterative cycle of \emph{design thinking} and
the phenomenon of \emph{design creativity} are hampered by the distribution of collaborating teams.

While the focus in research has been on technical artifacts of design (i.e., design notations and tools), there is only little work investigating design practices and cognition \cite{kan2012studing}.
This focus on technical aspects of design is problematic, as SE is a socio-technical endeavor, and 
further research controlling for human and social aspects is crucial for ensuring a successful engineering of software systems~\cite{williams2019methodology}.
In fact, Petre and Van Der Hoek \cite{petre2013software} argue that designing software is one of the least understood activities in which software developers engage. 

To close this gap, we conducted an exploratory multiple-case study\footnote{In this paper, we use the same data as in \cite{jolak2018does}, but with different objective and analysis.}, investigating (1) how design \emph{thinking} 
and \emph{design creativity} take place during software design in co-located and in distributed teams and (2) how remote collaboration affects the occurrence of \emph{creative events} between the participants.
To this end, we replicate a design study that was originally done by Petre and Van der Hoek \cite{petre2013software}.
This original study consisted of the recording and analysis of a design session that was held with multiple designers at a single location. In our replication, we used the same design-assignment, but asked designers that were distributed across geographically different locations to collaborate on creating a design. For this, the distributed developers used a tool that enables sketching in real-time and videoconferencing.
We then analyze the differences in between these design processes.
In particular, we qualitatively analyze how software developers switch between the problem and solution space of \emph{design thinking}, how alignment takes place, and how that affects the occurrence of \emph{creative events}~\cite{dorst01}.
Doing so, we aim to answer the following research questions.
\begin{itemize}
	\item \textbf{RQ1} How does distance affect the design thinking of software developers? 
	\item \textbf{RQ2} What challenges are encountered when collaboratively designing software at a distance?
	\item \textbf{RQ3} How does distance affect the design creativity of software developers? 
\end{itemize}  

Hence, the contributions of this paper are:
\begin{enumerate}
	\item We analyze the 
	\emph{design} process of software developers, leading to a better understanding of software design activities.
	\item We qualitatively compare how \emph{design thinking} differs between co-located and distributed setups, leading to an increased understanding of the effects of global SE on \emph{design thinking}.
	\item We analyze difficulties to distributed \emph{design thinking}, thus exposing areas of improvement in global SE projects.
	\item We qualitatively analyze the conversations between the participants to identify the occurrence of creative events in the locally and remotely collaborating teams as well as the discussions leading to these, which yields insights into the effects of remote collaboration on design creativity.
\end{enumerate}

This paper extends our findings reported in~\cite{jolak2020design} with 
a novel analysis of the effects on remote collaboration on creative events in software design and an in-depth discussion of related work on creative events in remote collaboration.

In summary, we find that distributed teams practice less \emph{design thinking} compared to co-located teams.
Furthermore, distributed teams focus more on solution space exploration and less on problem space exploration.
Our results indicate that a lack of awareness and 
common understanding between the remote collaborators cause this change in \emph{design thinking}.
However, this change of focus in design thinking does not seem to affect the occurrence of creative events.

The remainder of this paper is organized as follows:
We discuss related work in Section \ref{sec:related}, followed by a description of the multiple-case study design in Section \ref{sec:approach}.
We present the results in Section \ref{sec:results}, followed by a discussion in Section \ref{sec:discussion} and the threats to validity in Section \ref{sec:threats}.
We conclude and outline future work in Section \ref{sec:conclusion}. \section{Related Work}
\label{sec:related}

The results presented in this paper are related to \emph{Design Thinking}, \emph{Design Creativity}, and \emph{Global Software Engineering}, which are highlighted in the following.

\subsection{Design Thinking}

According to Kimbell~\cite{kimbell2011rethinking}, design thinking can be described as a cognitive style~\cite{cross2011design,dorst01}, a general theory of design~\cite{buchanan1992wicked}, or an organizational resource~\cite{martin2017design}. 
One school of thought considers \emph{design thinking} as an activity that the subject is aware of~\cite{dobrigkeit2019design,petre2019software}.
In contrast, several authors understand \emph{design thinking} as a theory that explains how subjects practice problem solving during a design task without necessarily being aware of it~\cite{dorst01,cross2011design,buchanan1992wicked}.
In this study, we consider the latter understanding of \emph{design thinking}.

\begin{figure}[t]
	\centering
	\includegraphics[width=0.5\columnwidth]{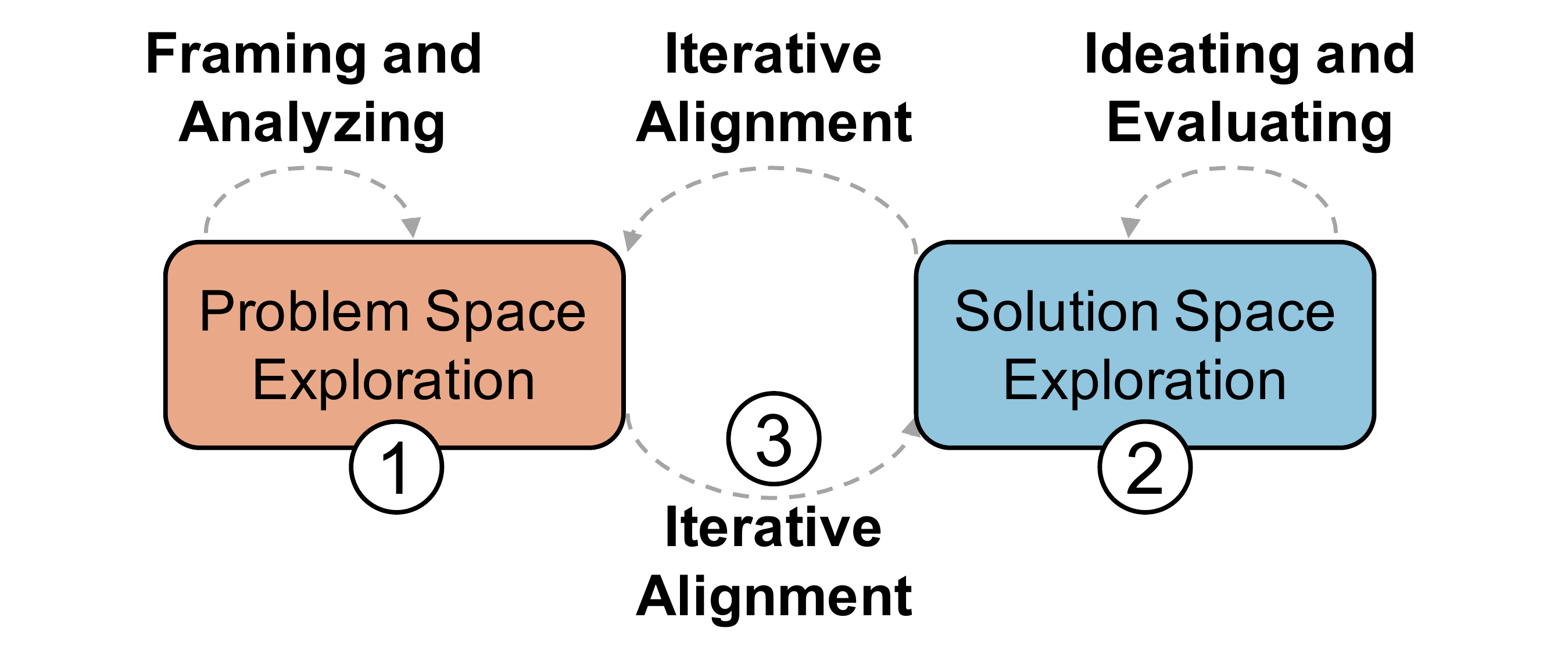}
	\caption{Problem solving with design thinking \cite{lindberg2011design}}
	\label{fig:designthinking}
\end{figure}

Lindberg et al. \cite{lindberg2011design} highlight that \emph{design thinking} fosters three main activities (see Figure \ref{fig:designthinking}):
\begin{enumerate}
	\item \textit{Exploration of the problem space}: by analyzing the problem space and framing the design problem; 
	\item \textit{Exploration of the solution space}:  by creatively devising and evaluating design solutions; and 
	\item \textit{Iterative alignment of both spaces}: by keeping the problem space in mind for refining and revising the chosen solutions.
\end{enumerate} 
Furthermore, Lindberg et al.~\cite{lindberg2011design} indicate that \emph{design thinking} can broaden the problem understanding and problem solving capabilities in IT development processes.
This is in line with Brooks \cite{brooks2010design}, who considers \emph{design thinking} an exciting new paradigm for dealing with problems in software and IT development.
Similar to Lindberg et al.~\cite{lindberg2011design}, Dorst and Cross~\cite{dorst01} find that a designer's understanding of the problem and solution space co-evolve in an iterative fashion, until the designer finds a \textit{bridge} that links concepts in the two spaces.
The authors write that "creative design involves a period of exploration in which problem and solution spaces are evolving and are unstable until (temporarily) fixed by an emergent bridge which identifies a problem-solution pairing. A creative event occurs as the moment of insight at which a problem-solution pair is framed"~\cite{dorst01}.
As a part of our study, we analyze these creative events in depth.

Petre et al.\cite{petre2019software} consider \emph{design} as a goal-driven activity to decide upon a plan for a novel change in a specific context.
This change, when realized, satisfies the contributing stakeholders.
They underline that design thinking is conducted in different social contexts and at all stages of software development. 
Moreover, the authors claim that developers with a `design-thinking mindset' perform contrasting design dialogues between problem and solution spaces, pragmatism and fitness-for-purpose, and across different levels of development focus and design cycles such as analysis, synthesis, and evaluation. 
Each of these dialogues provides a focus for design reasoning which helps to understand problems, manage complexity, and achieve enduring development success.

Dobrigkeit and de Paula~\cite{dobrigkeit2019design} investigate how design thinking can support software development and how it manifests itself during the development process.
By conducting a case study and interviews in a global IT company, Dobrigkeit and de Paula find that once trained in design thinking, developers find various ways to implement it throughout their projects even applying it to aspects of their surroundings such as the development process, team spaces and team work.

While, to the best of our knowledge, there are no studies that empirically assess the \emph{design thinking} phenomenon in SE, there are few studies focusing on the social and cognitive activities of developers when they collaboratively practice problem-solving activities. 
Jolak et al.~\cite{jolak2018does} study how distributed software designers communicate and collaboratively make design decisions.
They found that distance affects the quality of communication, and reduces the amount of design decisions that the developers take during their distributed collaboration.
Our multiple-case study uses the same data used in~\cite{jolak2018does}.
In contrast, in this study we focus on analyzing \emph{design thinking} and challenges to distributed collaboration. 

Similarly, Christiaans and Almendra~\cite{christiaans2010accessing} study how developers take decisions in software design.
They find that developers in general tend to prioritize their previous knowledge in problem solving, while neglecting a thorough analysis of information in order to take decisions along their processes.

For understanding the software design process, Baker and Van der Hoek~\cite{baker2010ideas} analyze how developers generate ideas, discuss subjects, and do design cycles (i.e., exploring a design aspect, make some progress, and then switch to a new design aspect).
They observe that the design process is highly incremental and developers repeatedly return to previously stated ideas and discussed subjects.
This observation is in line with how expert developers behave when practicing \emph{design thinking}, i.e., by exploring the problem space, the solution space, and then repeatedly jumping back and forth to align the two spaces.

Razavian et al.~\cite{razavian2016two} consider software design as a problem solving exercise.
They theorize that software design thinking requires two minds: a \emph{reasoning} mind that focuses on the process of logical design reasoning and a \emph{reflective thinking} mind that challenges the reasoning mind by asking reflective questions. 
Razavian et al. conduct multiple case studies to understand how reflections on reasoning and judgments influence software design thinking. 
They find that reflection improves the quality of software design discourse which, in turn, is considered as a foundation for a good design~\cite{dorst2011core}. 

In summary, \emph{design thinking} is a cognitive style that fosters problem and solution space exploration and alignment between the two spaces through so-called "bridges".
In software design, developers follow a process that is similar to \emph{design thinking}, thus making the combination of the two, software design and \emph{design thinking}, a relevant topic for investigation.

\subsection{Creativity in Design}
Dorst and Cross~\cite{dorst01} characterize creativity in the design process as a phenomenon that occurs when designers have sudden, significant insights leading to the emergence of a key design concept or decision.  
They consider studies of creativity in design as necessary to develop a better understanding of how creative design occurs.
Accordingly, they conduct protocol studies of nine experienced designers working on a small design assignment in order to observe creativity in the design. 
Activities such as defining and framing of design problem, ideation of original decisions, and co-evolution of the problem space and solution space (driven by surprising information) are reported to be key aspects of creativity. 
Moreover, different factors are observed to influence the design creativity, such as designer's own design goals and available design time.

Kruger and Cross~\cite{kruger2006solution} analyze different cognitive design strategies employed by designers, and relate these strategies to design quality and creativity.
The different analyzed design strategies are: problem driven, information driven, solution driven, and knowledge driven design strategies. 
The results indicate that designers using a solution driven design strategy tend to have a higher creativity, but lower overall solution quality. 
Furthermore, designers using a problem driven design strategy tend to have the best results in both design quality and creativity. 

Wiltschnig et al.~\cite{wiltschnig2013collaborative} examine real-world design data from team-based design and development meetings. 
They analyze the design co-evolution within a collaborative context where several designs collaboratively create a design solution, and examine whether the design problem and solution co-evolution concept captures collaborative creativity.
The results show that design co-evolution iterative episodes are related with creative activities, such as analogizing and mental simulation which are critical for overcoming moments of uncertainty and for facilitating problem understanding and solution generation~\cite{ahmed2009situ}. 
Hence, the results support the view that design co-evolution is a driver of creativity in collaborative design.

Creativity involves both divergent and convergent thinking~\cite{gabora2010revenge}.
Divergent thinking moves the thought away to consider different aspects that can foster new ideas and creative solutions. 
Convergent thinking brings together information and knowledge in order to solve problems.
Maiden and Robertson~\cite{maiden2005integrating} explore the integration of creativity as well as divergent and convergent thinking into the requirement engineering process.
In particular, they study the effectiveness of a number of techniques for supporting the types creativity as identified by Boden~\cite{boden2004creative}: exploratory, combinational, and
transformational creativity. 
Exploratory creativity enables seeing possibilities and investigating new idea.
Combinational creativity enables the alignment of ideas and requires the ability to find links between concepts.
Transformational creativity involves the transformation of some dimensions of the thinking space, so that new ideas can be generated.
The results of Maiden and Robertson show that brainstorming is more effective than analogical reasoning for exploratory creativity.
Storyboard development is found effective for combinational creativity.
Moreover, results revealed that removing constraints lead to the generation of more ideas, and thus supporting transformational creativity.  

Mohanani et al.~\cite{mohanani2019requirements} study the effect of presenting desiderata as ideas, requirements or prioritized requirements on design creativity. 
They find that using desiderata framed as requirements or prioritized requirements lead to the creation of designs that are less original, but more practical than the designs created by using desiderata framed as ideas.
Accordingly, it is suggested that more formal and structured presentations of desiderata are less appropriate where a creative solution is desired.

Nguyen and Shanks~\cite{nguyen2009framework} provide a framework for understanding creativity in requirement engineering.
They state that a creative product is characterized by being novel, original, and useful.
Moreover, they state that surprisingness is often associated with creative products. 
Considering creative processes, an inspirationalist view is discussed. 
According to this view, a creative process is characterized by moments of preparation, incubation, illumination, and verification and expression of insights. 
These insights occur at the ``Aha!'' moment, when a long-sought idea or solution suddenly appears at the conscious level.

\subsection{Global Software Engineering}

Global software engineering (GSE) is the practice of engineering software systems across geographical, socio-cultural, and temporal boundaries~\cite{herbsleb2007global}.
Software organizations opt for globalizing their projects mainly to maximize business profits by taking advantage of low development cost and time, achieving a high percentage of productivity, accessing a skillful workforce and using innovative concepts~\cite{o2009benefits}. 
Also, in 2020 the COVID-19 epidemic has forced many software development teams to work in a distributed manner - even while their team-members still live in the same city.
However, these organizations often face numerous challenges, including poor quality of globally developed software~\cite{khan2019gsepim}.

Herbsleb~\cite{herbsleb2007global} studies the impact of geographic distance on distributed collaboration.
He suggests that co-location facilitates communication since software developers explicitly know who is working and what is happening in the working place.
The suggestion of Herbsleb is in line with Damian et al.~\cite{damian2007collaboration}, who observe that geographic distance hampers awareness of remote collaborating teams participating in GSE.

Besides geographic distances, communication gaps can be caused by other distances.
Bjarnason et al.~\cite{bjarnason2016theory} present a theory on different distances and their influence on communication and coordination in software development projects.
The authors suggest that some of these distances can be shortened by following certain practices such as:
\begin{itemize}
    \item involving roles from different disciplines to perform an SE activity, 
    \item reviewing documentation and artifacts, or 
    \item performing incremental SE.
\end{itemize}

Of particular interest to our study are communication gaps caused by social distances, since these distances are often amplified by geographic distance and since they cannot easily be mitigated by technological solutions.
For instance, as part of a substantial body of work on culture, the Hofstede culture distances~\cite{hofstede2001culture} are a well-known theory aiming to explain how different cultural backgrounds developed in families, schools, and organizations introduce differences in thinking and social actions.
Geographic distribution, e.g., due to outsourcing, often lead to a more diverse mix of cultures which can give rise to communication gaps.
For instance, Lehmann-Willenbrock et al. \cite{lehmann2014observing} observe that culture can introduce differences in the behaviour of distributed collaborating teams.
In addition to the culture dimension, Bj\o{}rn et al.~\cite{bjorn2014does} find that geographic distance raises more social challenges which are considered critical obstacles for successful remote collaboration. 
Overall, we see that more in depth studies of the activities and behavior of distributed developers are needed~\cite{jolak2019position}.
This should result in a better understanding of how to account for technological and social challenges caused by geographic distance.

The design assignment that we use in our study is intentionally formulated to trigger design thinking and reasoning. 
Informal notations and whiteboards are often used during collaborative design ideation and reasoning to explore problems and externalize design solutions~\cite{cherubini2007let}.
Whiteboards do not constrain the modeling notation that can be used and, thus, support informal modeling and design thinking.
Therefore, in our study we use whiteboards (standard whiteboards in the co-located case and interactive whiteboards in the distributed case) to better study the design thinking phenomenon. 

Dekel~\cite{dekel2005supporting} study co-located software design meetings in order to observe the activities of developers and outline requirements for tools supporting distributed software design.    
For distributed software design, Dekel suggests that the design tools should mainly support the creation of informal notations to capture ideas while brainstorming.
There are several tools that support distributed software design and the creation of informal notations, such as the Software Design Board~\cite{wu2004software}, Calico~\cite{mangano2010software}, Metaglue~\cite{hammond2002agent}, and OctoUML~\cite{jolak2017octouml} (a tool that we developed).
In this study, we use interactive whiteboards with a simplified version of OctoUML to support and explore distributed software design. 
More details are provided in the next section.

 \section{Case Study Design}
\label{sec:approach}
The intention of this study is to explore the \emph{design thinking} and \emph{design creativity} of co-located and distributed software developers.
Moreover, we seek to identify challenges and impediments that could hinder collaborative distributed \emph{design thinking}. 
We aim to seek insights and generate hypotheses for future research by observing the process and outcome of the \emph{design thinking} and \emph{creativity} phenomena.
But as it is hard to study \emph{design thinking} and \emph{creativity} in isolation and separate it from its context, we chose a case study design, where the boundary between the phenomenon and its real-life context cannot be clearly specified \cite{yin2017case}.
Our case study is an exploratory and inductive multiple-case study \cite{runeson2012case}.

\subsection{Cases and Units of Analysis}
\begin{figure}[!t]
	\centering
	\includegraphics[width=0.5\columnwidth]{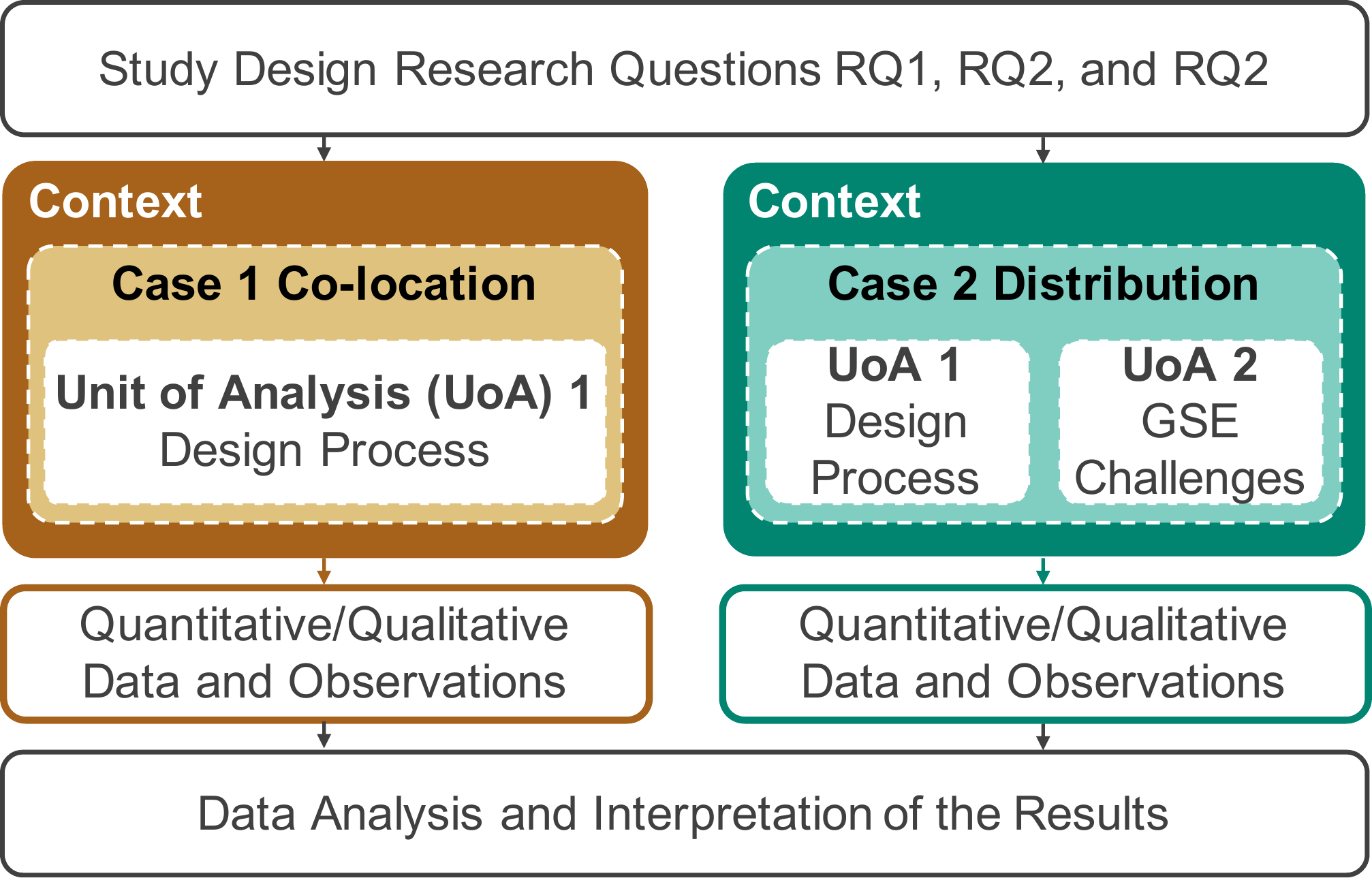}
	\caption{Multiple-case study design featuring the two use cases of co-located and distributed software design}
	\label{fig:studydesign}
\end{figure}

Figure \ref{fig:studydesign} shows the design of our multiple-case study, which serves to examine two cases:

\begin{itemize}
	\item \textbf{Case 1} (\textit{Co-location}): Collaborative co-located designing of software architecture using a whiteboard. This is a single case \cite{yin2017case} with one Unit of Analysis (UoA): \emph{Design Process}. Here, we analyze the problem-solving cognitive style of developers and, in particular, focus on \textit{Design Thinking} and \emph{Design Creativity} phenomena. 
	\item \textbf{Case 2} (\textit{Distribution}): Collaborative distributed designing of software architecture using an interactive whiteboard. This is an embedded case \cite{yin2017case} with two UoA: \emph{Design Process} and \textit{GSE Challenges}. In this case, we analyze the challenges that hinder collaborative distributed designing. 
\end{itemize}

\subsection{Theoretical Framework}
While our case study is exploratory in nature, even this kind of case study should have a  foundation consisting of theories, if possible, and propositions/hypotheses \cite{runeson2012case,yin2017case}.
Next, we detail the theoretical framework that we employ in this multiple-case study.

In this study, we analyze the software design process and reason about how individuals solve problems. 
First, we make use of the \emph{design thinking} phenomenon and analyze the interactions of individuals to  
the three \emph{design thinking} activities described by Lindberg et al.~\cite{lindberg2011design}: Exploration of the problem space, of the solution space, and iterative alignment between the two.
While developers can be explicitly trained in \emph{design thinking}, we follow the line of research that assumes that even untrained developers perform \emph{design thinking}.
Second, we analyze the design discussions to explore how creative design occurs i.e., understand the events, activities, or behaviors that contribute to design creativity~\cite{dorst01}. 
The two cases are selected to predict possible contrasting results on 
the \emph{design process} by altering one condition: the geographic distance (co-location vs. distribution).

In addition to the \emph{design thinking} and \emph{creativity} phenomena, we analyze our data with respect to Bjarnason's theory on distances in SE \cite{bjarnason2016theory}.
Using these distances, we aim to reason about the differences between the two cases, as well as the perceived challenges of being distributed (UoA 2 in Case 2).

The related work on GSE, discussed in Section~\ref{sec:related}, shows that remote collaboration can hinder effective communication.
We relate this overall conclusion to our RQs by making the following three propositions, which then guide our case study design:

\begin{itemize}
	\item \textbf{Proposition A1}: We propose that geographic distance (co-location vs. distribution) does not affect \emph{design thinking}. 
	\emph{If} the \emph{design thinking} of the two cases varies considerably, \emph{then} this indicates that geographic distance affects \emph{design thinking}. This would encourage deeper investigations of the effect of distance on \emph{design thinking}.
	\item \textbf{Proposition A2}: We propose that geographic distance (co-location vs. distribution) does not affect \emph{design creativity}. 
	\emph{If} the \emph{design creativity} of the two cases varies considerably, \emph{then} this indicates that geographic distance affects \emph{design creativity}. This would encourage deeper investigations of the effect of distance on \emph{design creativity}.
	\item \textbf{Proposition B}: We propose, based on the findings in \cite{olson2000distance,herbsleb2007global}, that poor GSE tool-support and difference imposed by social factors are the most frequently reported challenges that affect GSE.
	\emph{If} the perceived challenges in our study differ from those reported in related work, \emph{then} this indicates that additional factors in our study setup influence the perceived challenges. Moreover, this would encourage further investigation of the perceived challenges and confounding factors. 
\end{itemize}

\subsection{Context}
\begin{figure*}[t!]
	\centering
	\includegraphics[width=0.98\textwidth]{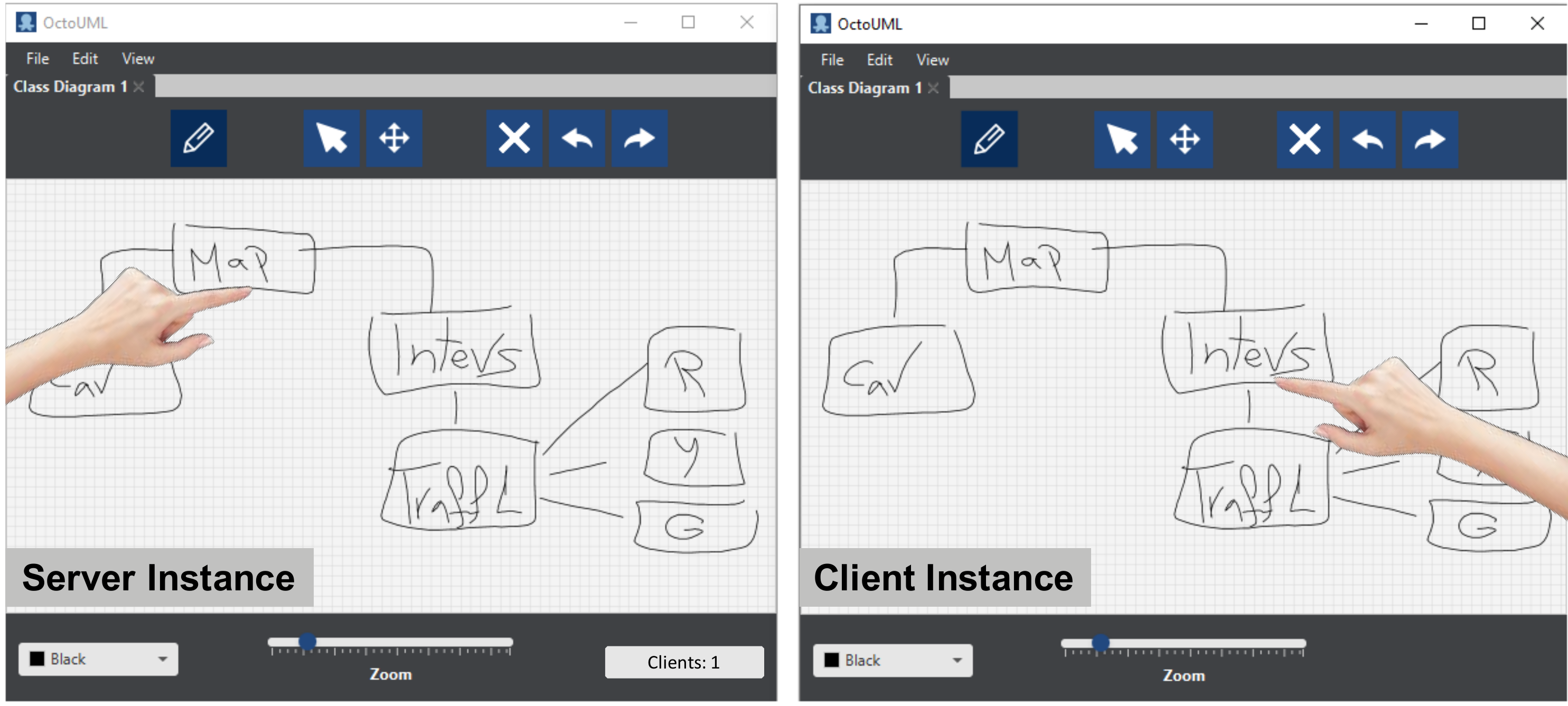}  
	\caption{The main views of the server and client instances of OctoUML}
	\label{fig:octoUML}
\end{figure*}

For both cases, we used a software design challenge, in which designers are asked to design the signal timing at a road intersection with the goal that traffic is flowing in a fluid manner and waiting times are minimized~\cite{petre2013software}. 
The challenge documentation is available online\footnote{Challenge Documentation: \url{https://www.ics.uci.edu/design-workshop}.}.
The designers are asked to create the architecture of a simulator enabling its users to investigate the effect of different signal timing on the traffic flow. 
The challenge is of realistic size and complexity, and focuses on four functional requirements:
\begin{enumerate} 
\item Users can create a visual map of intersected roads of varying length.
\item Users can describe the behavior of the traffic lights at each of the intersections, such that combinations of individual signals that would result in crashes are prohibited.
\item Users can simulate traffic flows on the map, and the resulting traffic levels are conveyed visually.
\item Users can change the traffic density per road.
\end{enumerate}

Prior to starting the challenge, developers were informed that:
\begin{itemize}
    \item their design will be evaluated primarily on the basis of its elegance and clarity, and
    \item they should focus on the interaction that the users will have with the system, including the basic appearance of the program, and on the important design decisions that form the foundation of the implementation.
\end{itemize}

Developers were given a printed copy of the design challenge and had a maximum of $2$ hours to work on their design. 
The developers were given no additional instructions, other than to use the whiteboard for any writing or drawing that they wished to perform.
After completing their work, the developers were given $10$ minutes to collect their thoughts and briefly explain the design.

For Case 1, we used the data set provided by Petre and Van der Hoek\cite{petre2013software}, who performed the study with three (co-located) teams of two professional software developers each.
The developers in these teams were considered expert designers (i.e., their companies would trust them in solving key software design problems) and had on average 19 years of experience.

For Case 2, we recruited three teams of two professional software developers to work on the same design challenge, but from two different geographic locations: Aachen, Germany and Gothenburg, Sweden. 
The developers in our study had between three and seven years of professional software development experience in automotive and networking domains, and were selected based on convenience sampling.

Instead of a regular whiteboard, we used interactive whiteboards with a simplified version of OctoUML\footnote{OctoUML Website: \url{http://rodijolak.com/\#octouml}}, an open source software design environment \cite{jolak2017octouml} supporting remote collaborative design sessions between geographically distributed developers.
To collaborate remotely, one developer creates a \emph{server} instance of OctoUML.
Other remote developers can connect as \emph{client}s using TCP/IP protocol. 
Figure~\ref{fig:octoUML} shows the main views of the server and client instances of OctoUML.
Developers can simultaneously sketch on the shared canvas of OctoUML using special styluses or by using their fingers, as interactive whiteboards support touch-input.
While OctoUML has some UML capabilities, like creating class shapes, we removed those in the study to make OctoUML resemble a regular whiteboard as closely as possible and, thus, mitigate the impact of other factors than the geographic distance on the UoA 1 (i.e., design process) of the two cases. 
Similarly, choosing another commercial remote collaboration tool might have introduced other collaboration features that would have changed the experience compared to a regular whiteboard.
Hence, we opted to choose OctoUML since we could customize it for the study's purpose.
The interactive whiteboards were connected to computers providing video conferencing (via Skype\footnote{Skype Website: \url{https://www.skype.com/en}}) between the two locations. 

\begin{figure*}[!h]
	\centering
	\includegraphics[width=0.93\textwidth]{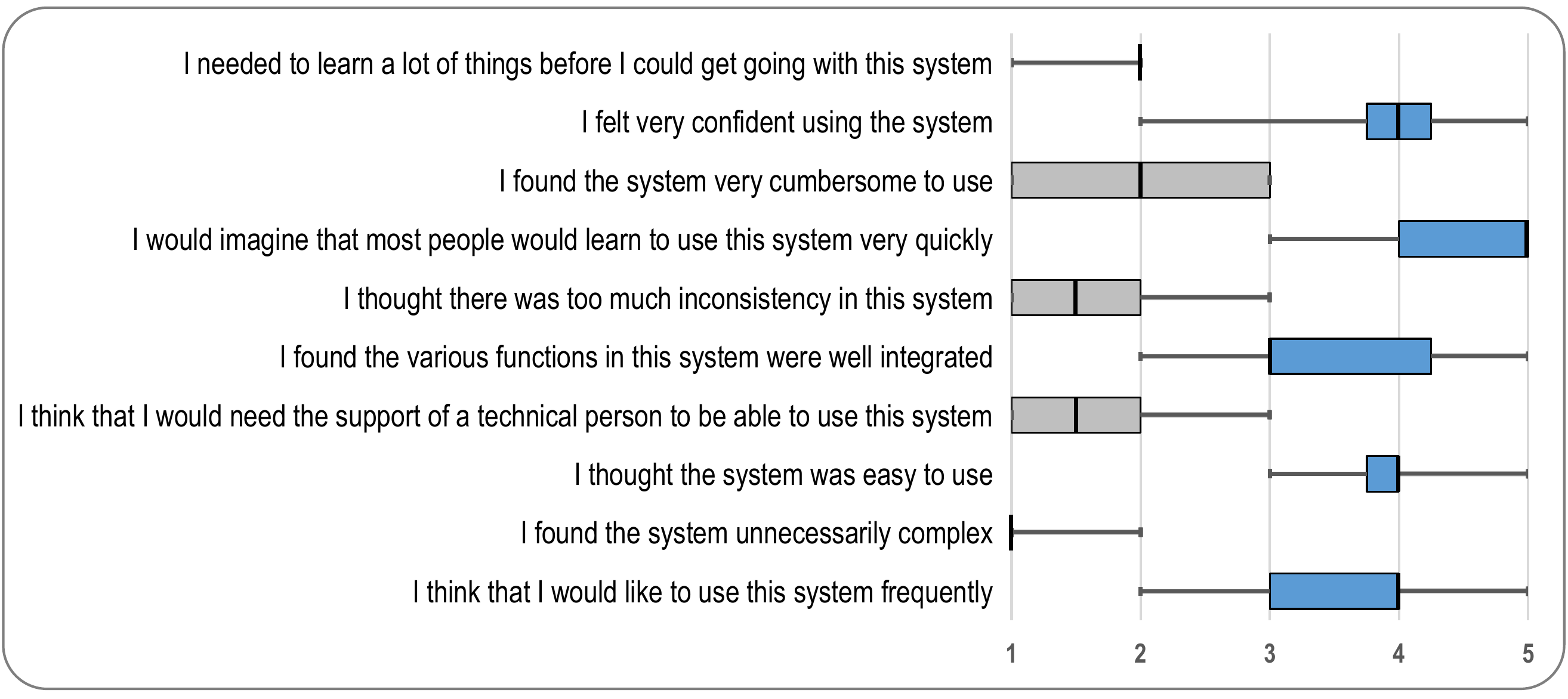}
	\caption{The perceived usability of OctoUML}
	\label{fig:usability}
\end{figure*}

Immediately after each distributed design session, we asked the developers to the evaluate the usability of OctoUML.
The reason is to understand to which extent the usability of OctoUML did impact the work of the distributed developers.
In particular, we asked the developers to answer the System Usability Scale (SUS) questionnaire.
The System Usability Scale is an easy, standard way of evaluating the usability of a system~\cite{brooke2013sus}. 
It is a form containing ten statements, and users provide their feedback on a 5-point scale (1 is ``strongly disagree'' and 5 is ``strongly agree''). 
SUS effectively differentiates between usable and unusable systems by giving a measure of the perceived usability of a system. 
It can be used on small sample sizes and be fairly confident of getting a good usability assessment~\cite{tullis2004comparison}.
Figure \ref{fig:usability} shows the results of SUS evaluation.
Overall, we observe that the perceptions regarding the usability of OctoUML were positive. 
Regarding the SUS score, OctoUML received an average SUS score of $74.17 \pm 5.63$, which can be interpreted as a grade of $B-$ (i.e., a good usability score) according to \cite{sauro2011practical}.

\subsection{Data Collection and Analysis}
In both cases, design sessions were recorded with a video camera that was positioned to capture the whiteboard and the developers at work.
In Case 2, we recorded two videos per session, one each for the two geographic locations.
We then used verbatim transcriptions for subsequent analysis (for Case 1, the transcriptions were available online\footnote{Case 1 Transcriptions: \url{https://www.ics.uci.edu/design-workshop/videos.html}} ).

\vspace{5pt}
\subsubsection{Design Thinking}
\begin{figure*}[!ht]
	\centering
	\includegraphics[width=.93\textwidth]{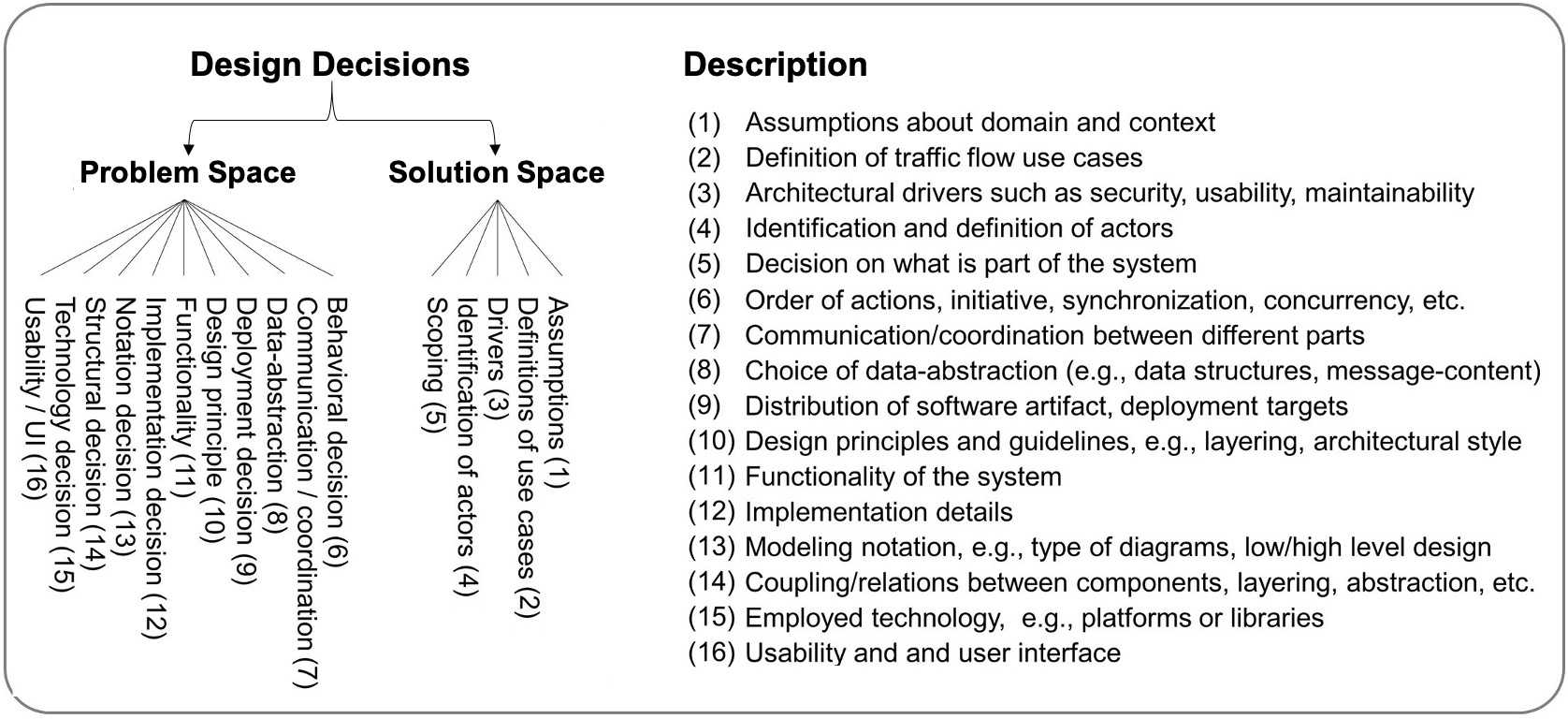}  
	\caption{Classification schema of design decisions into problem space and solution space decisions (based on \cite{weinreich2015expert})}
	\label{fig:designD}
\end{figure*}

We analyzed the transcriptions of approximately $10$ hours of design activity by six pairs of professional software developers, and performed a manual coding of more than $2000$ conversation dialogues.  
We created a coding schema (see Figure \ref{fig:designD}) to capture design decisions from the problem space (traffic flow) and solution space (SE). 
This schema is based on the design-reasoning decisions of Weinreich et al. \cite{weinreich2015expert}.
Dialogues in the transcriptions were then assigned codes (using NVivo\footnote{NVivo Website: \url{https://www.qsrinternational.com/nvivo}}).
If multiple codes were assigned per dialogue, we ordered them by the time they occurred in the transcription.

Two coders ensured reliability, following established advice on selecting and reporting intraclass correlation coefficients~\cite{koo2016guideline}.

We performed two-way mixed Intraclass Correlation Coefficient (ICC (3,k)) tests with $95$\% confidence interval on $11$\% of the data.
The ICC value is \emph{$0.84$}, 
which is considered a good reliability~\cite{koo2016guideline}.  
After that, the two coders discussed and aligned the differences in their coding.
Finally, the two coders collaboratively continued to code the rest of the data i.e., $89$\% of the data.
Based on the frequency of occurring codes and their order, we derive the \emph{design thinking} graphs reported and discussed in Section \ref{sec:results}. 
Essentially, we derive the frequency of each \emph{design thinking} phase by counting (see Figure \ref{fig:designD} as a reference): 
\begin{itemize}
    \item how frequent the developers explored the problem space (codes 1-5),
    \item how frequent the developers explored the solution space (codes 6-16), and
    \item how frequent the developers iteratively align the two spaces (subsequent codes changing from problem- to solution space, or vice versa).
\end{itemize}

\subsubsection{Design Creativity}
In addition to this quantitative summary of the codes, we performed a qualitative analysis of \emph{creative events}~\cite{dorst01} in the interactions between each pair of designers.
A creative event in \emph{design} is the "moment of insight" at which "an emergent bridge which identifies a problem-solution pairing [..] is framed" \cite{dorst01}.
Similarly, Nguyen and Shanks~\cite{nguyen2009framework} describe moments of insights, or ``Aha!'' moments.
Each transcript was re-read by at least two of the authors, trying to identify creative events based on the following descriptions derived from Dorst and Cross \cite{dorst01}:
\begin{enumerate}
    \item Subjects understand a connection that they were unaware of beforehand. This connection simplifies an issue they were dealing with so far, resulting in a solution to a part of the assignment/task. OR
    \item Subjects identify an issue they were unaware of before, leading to a complication of the overall task. OR
    \item Subjects find an appealing simple solution to an issue. This greatly simplifies the issue or the overall task. OR
    \item In terms of keywords, the subjects show sudden realization (e.g., "Aha", "Now I understand it", "Wait a moment, we haven't thought about..."). OR
    \item The subjects feel that they have understood or uncovered the underlying problem behind or the difficulty with the task.
\end{enumerate}

Keywords alone do not clearly identify a creative event, as they are context-dependent (e.g., a matter of fact "Now I understand what you are saying" vs. an emotional "Now I finally understood the assignment").
Therefore, this task is to some extent subjective.
Dorst and Cross however state that creative events are generally "highly emotional steps"~\cite{dorst01}.
Since we did not find any events that would correspond to this strict classification of creative events as "highly emotional step", we relaxed the coding guidelines to also include general realizations that were not emotional in nature.
This is further discussed in Section~\ref{sec:discRQ3}.

After identifying all creative events, the authors processing the transcripts coded the discussion before and after the creative event using an open, descriptive coding approach \cite{saldana15}.
We then used the resulting high-level codes to summarize how the subjects reached the creative events, and what distinguished different approaches.
Afterward, we assigned two labels to each event.
First, we identified whether the creative event was concerned with the problem space, the solution space, or a bridge between the two spaces.
This links our analysis of creative events to the \emph{design thinking} phenomenon and allows us to compare and discuss the findings of both analyses.
Second, we classified whether the discussion before reaching the creative event was convergent or divergent in nature.
The presence of a combination of convergent and divergent discussions is believed to be an indicator of design creativity \cite{goldschmidt16}.
The two discussion modes are related to two thinking modes from a neurological point of view \cite{gabora2010revenge}.
Divergent discussions are characterized by "defocused attention" \cite{gabora2010revenge}.
We coded discussions as divergent if the topic being discussed shifted multiple times in the time leading up to a creative event.
In contrast, convergent discussions are characterized by "focused attention" \cite{gabora2010revenge}, which we assigned to discussions that essentially focused on a single topic or concept.

We plotted the resulting creative events with their associated labels on a time line, and discussed differences in frequencies, location, and labeling of creative events between different teams.

\subsubsection{Challenges to Distributed Design}
In each geographically distributed site (Sweden/Germany), one supervisor attended the design sessions to observe the design process and note observed challenges.
In addition, after each distributed design session, we asked the developers to indicate, and elaborate on, eventual challenges to their distributed collaborative design experience via an online form (self-evaluations) using the following two open questions:
\begin{itemize}
    \item Q1. What was challenging in this experience and what was missing in your opinion?
    \item Q2. Did you recognize challenges due to being geographically distributed? If you did, which challenges did you find?
\end{itemize}
Collecting data on GSE challenges from two sources (perceptions of the distributed developers and our observations of the design sessions) allowed us to triangulate the data on the experienced challenges.

 \section{Results}
\label{sec:results}

In this section we present the results of this study based on our three RQs. A detailed discussion of those results is provided in Section~\ref{sec:discussion}.

\subsection{RQ1: Co-located vs. Distributed Design Thinking}
\begin{figure*}[!ht]
    \vspace{18pt}
    \centering
    \captionsetup{justification=centering}
    \begin{subfigure}[b]{0.3\textwidth}
        \includegraphics[width=\textwidth]{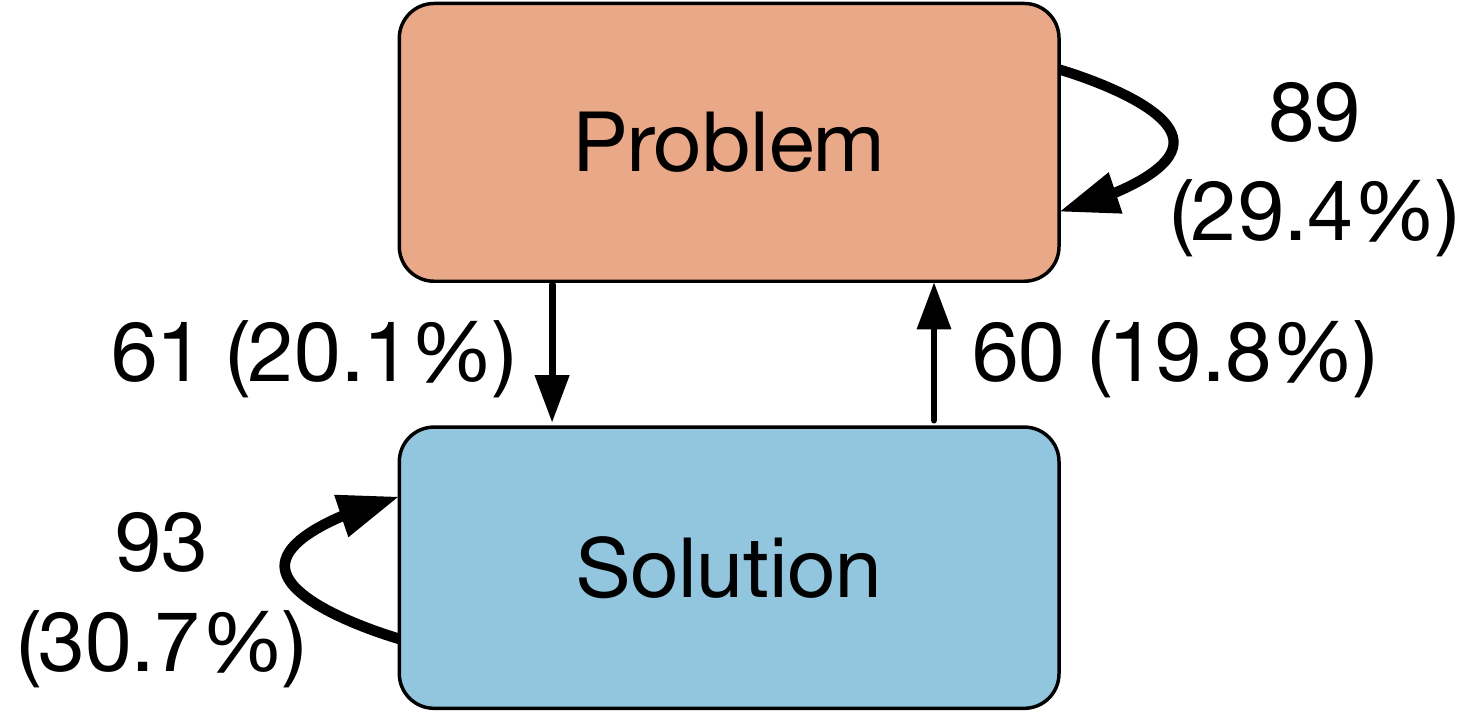}
        \caption{CT1}
        \label{fig:CT1}
    \end{subfigure}
    \hfill
    \begin{subfigure}[b]{0.3\textwidth}
        \includegraphics[width=\textwidth]{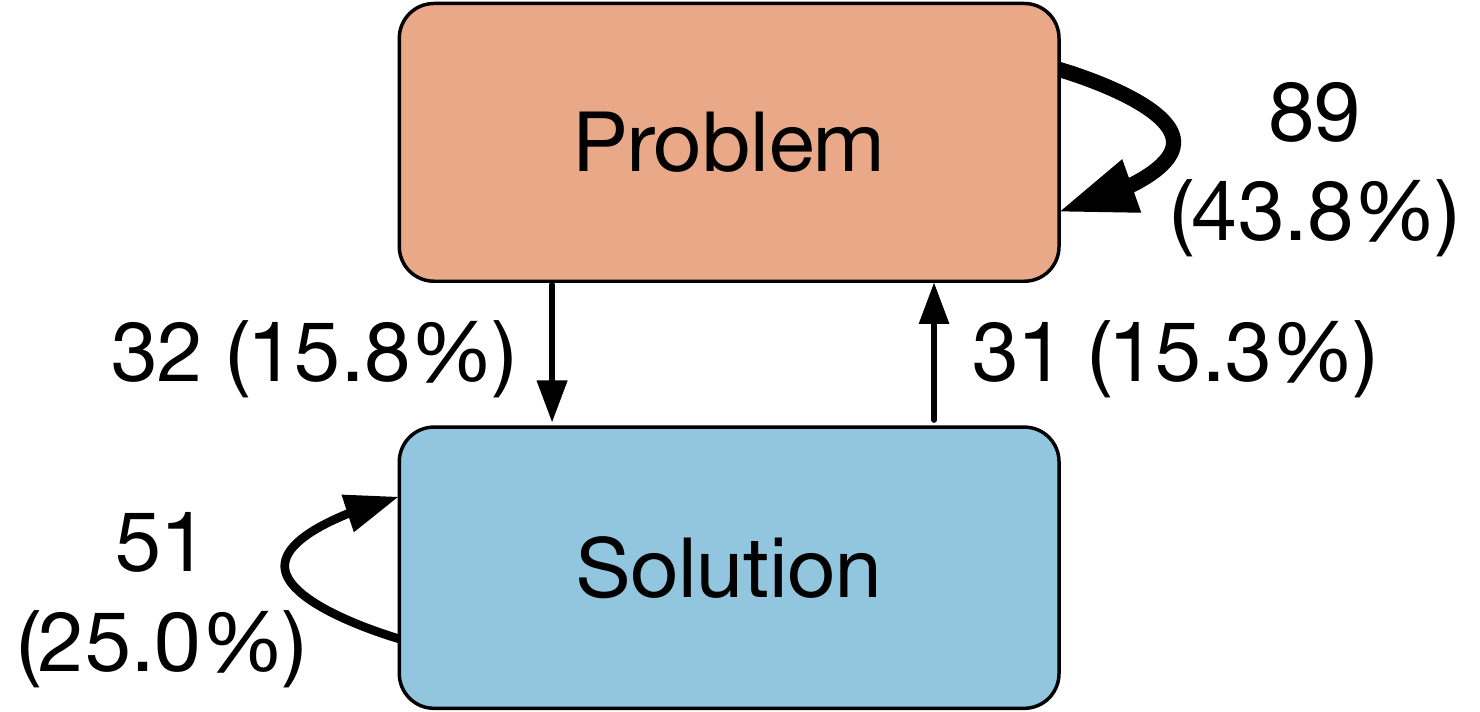}
        \caption{CT2}
        \label{fig:CT2}
    \end{subfigure}
    \hfill
    \begin{subfigure}[b]{0.3\textwidth}
        \includegraphics[width=\textwidth]{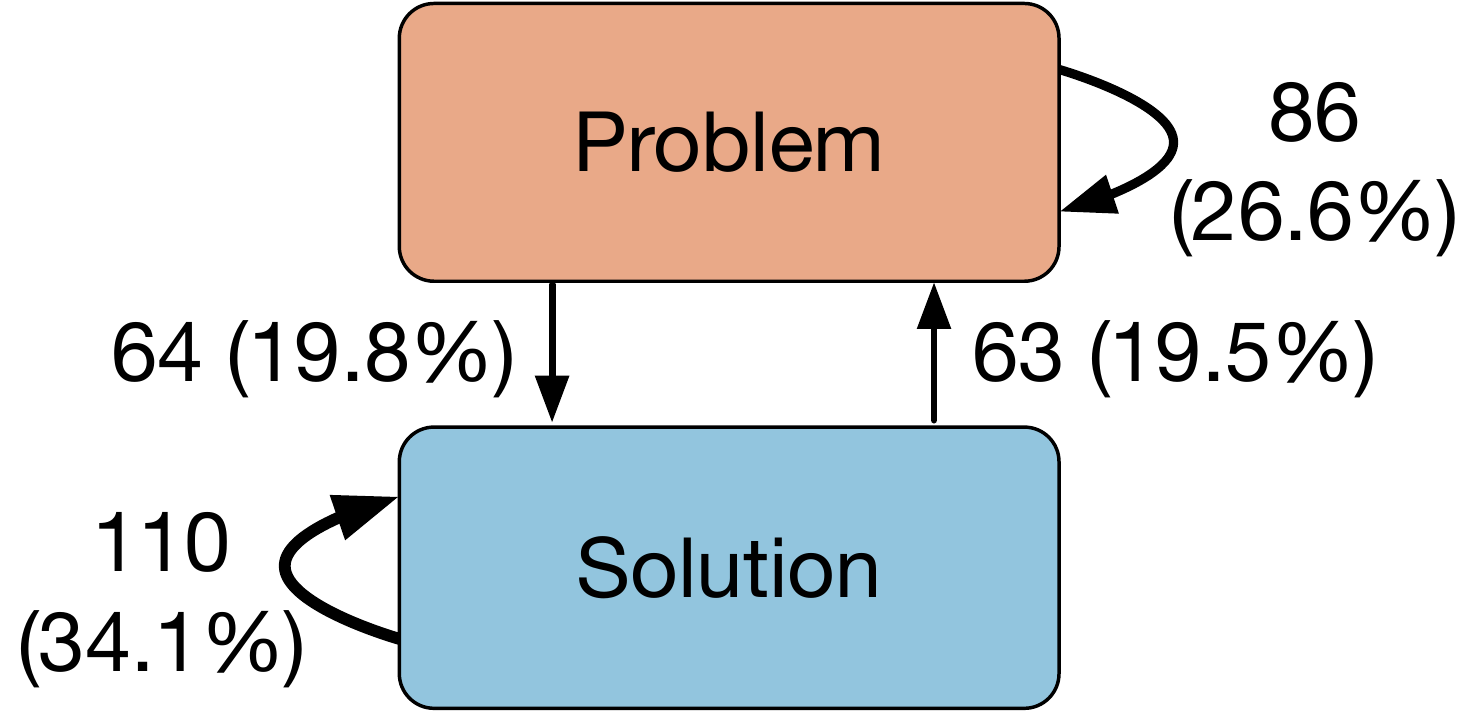}
        \caption{CT3}
        \label{fig:CT3}
    \end{subfigure}
    \caption{The \emph{design thinking} of Co-located Teams CT1, CT2, and CT3 (Case 1))}
    \vspace{1cm}
    \label{fig:colocatedRes}
\end{figure*}

\begin{figure*}[!ht]
    \centering
    \captionsetup{justification=centering}
    \begin{subfigure}[b]{0.3\textwidth}
        \includegraphics[width=\textwidth]{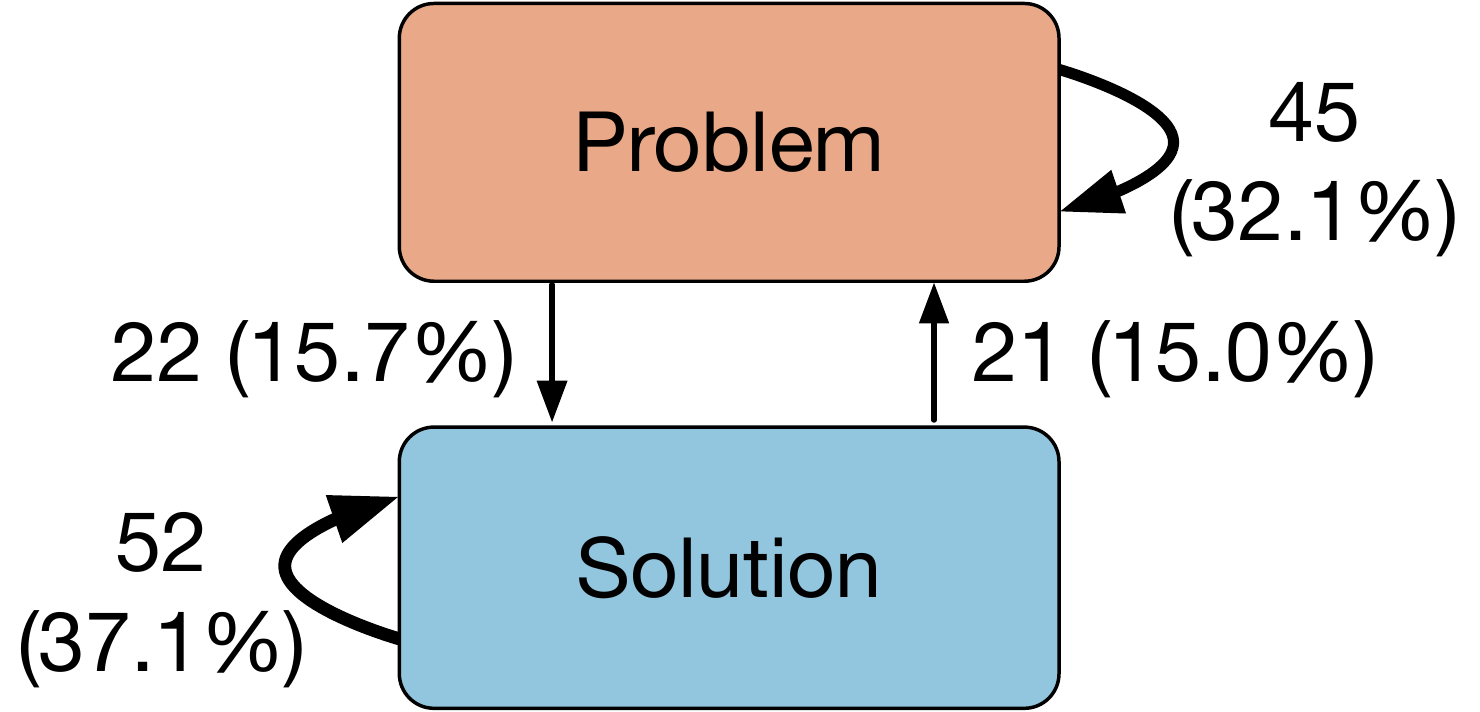}
        \caption{DT1}
        \label{fig:DT1}
    \end{subfigure}
    \hfill
    \begin{subfigure}[b]{0.3\textwidth}
        \includegraphics[width=\textwidth]{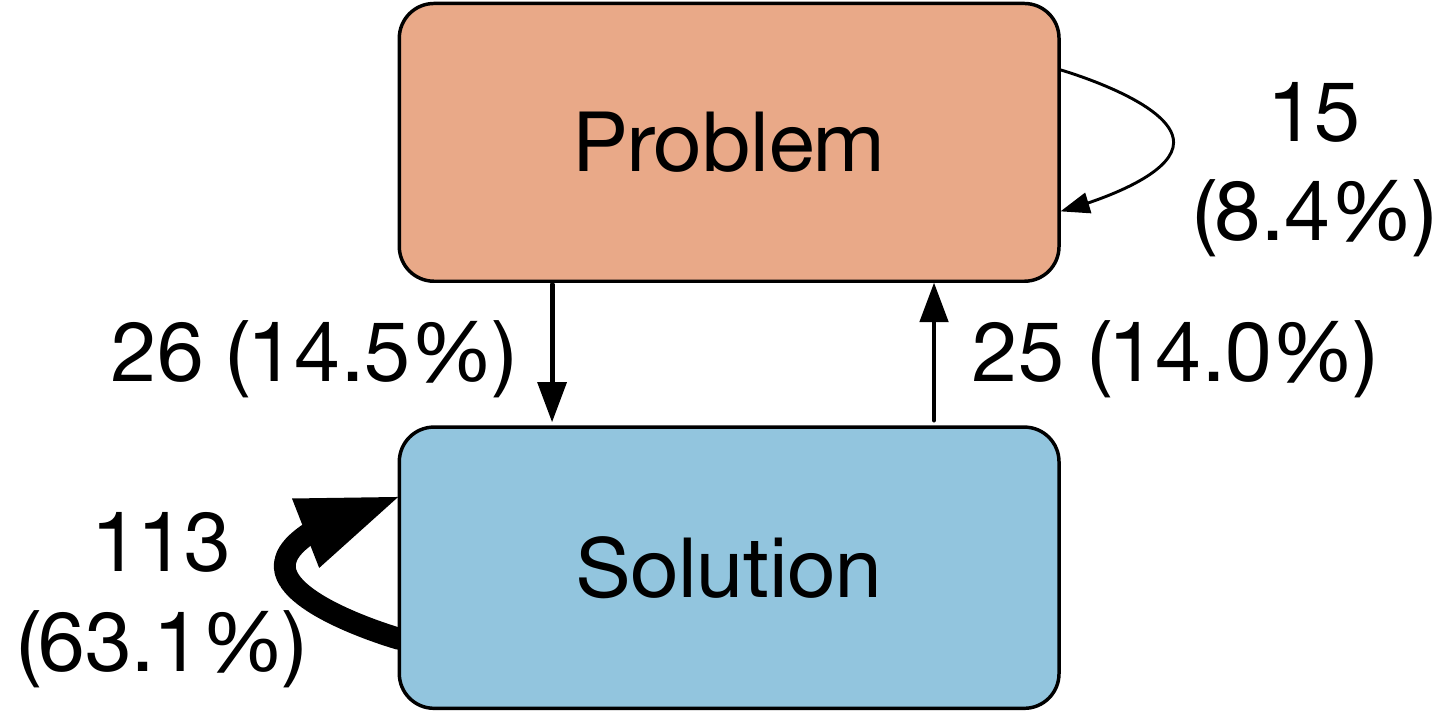}
        \caption{DT2}
        \label{fig:DT2}
    \end{subfigure}
    \hfill
    \begin{subfigure}[b]{0.3\textwidth}
        \includegraphics[width=\textwidth]{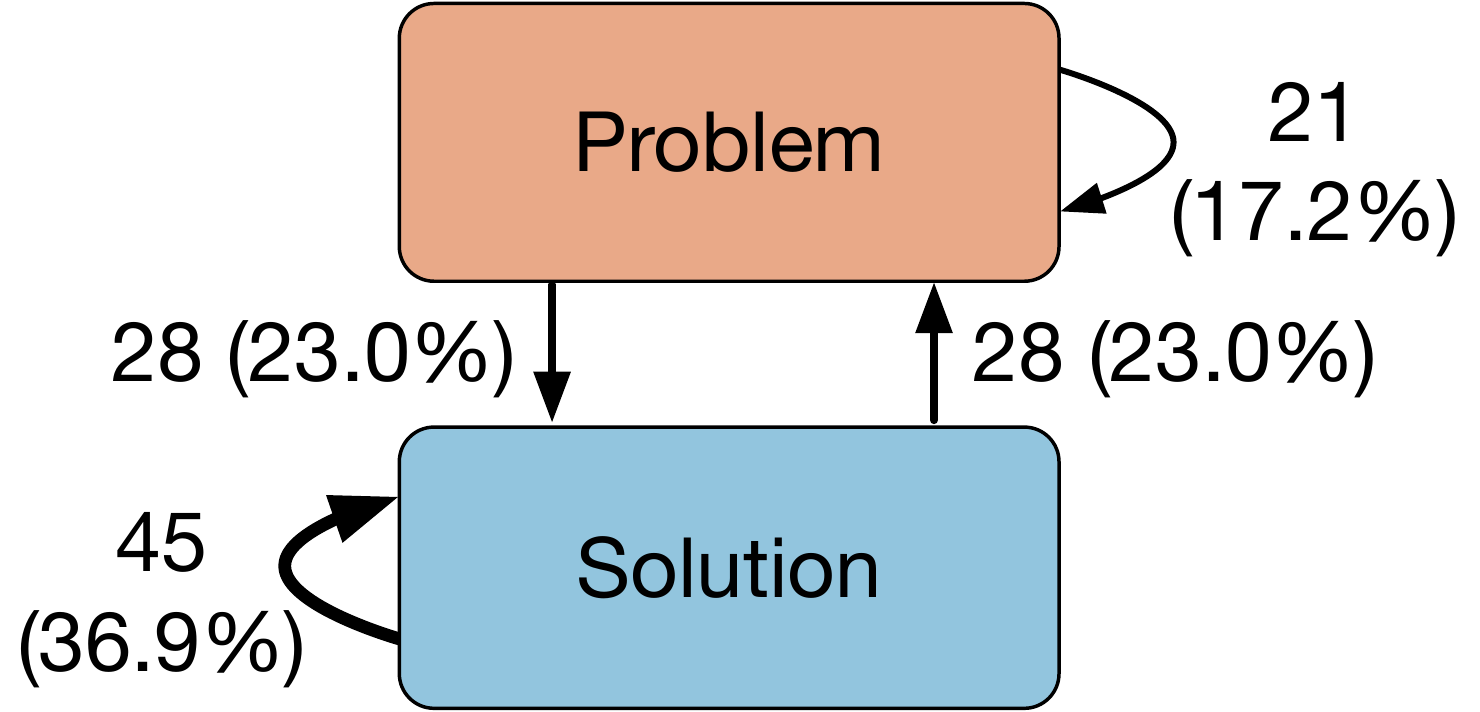}
        \caption{DT3}
        \label{fig:DT3}
    \end{subfigure}
    \caption{The \emph{design thinking} of the Distributed Teams DT1, DT2, and DT3 (Case 2)}
    \label{fig:distributedRes}
\end{figure*}

Figures \ref{fig:colocatedRes} and \ref{fig:distributedRes} show the \emph{design thinking} processes of the co-located and distributed teams, respectively.
The numbers in the figures represent the absolute and relative frequencies of (i) problem space
exploration, (ii) solution space exploration, and (iii) alignment of the two spaces, that are practiced by each team in Cases 1 and 2.

Looking at absolute frequencies, we observe a higher number of \emph{design thinking} interactions in the co-located teams (CT1: $303$, CT2: $203$, CT3: $323$ interactions) than in the distributed teams (DT1: $140$, DT2: $179$, DT3: $122$ interactions).
Furthermore, we notice that the teams in Case 1 did more problem space exploration and more alignment between the problem space and solution space than the teams in Case 2.

In terms of relative frequency, we see that all teams in Case 2 have a larger percentage of solution space exploration than any of the teams in Case 1.
Specifically, the two developers in DT2 were more solution oriented (i.e., 63.1\% of their \emph{design thinking} was into the solution space), and did more solution space exploration than any other team, co-located or distributed.
Focusing only on Case 2, we notice that all teams explored the problem space (DT1: 32.1\%, DT2: 8.4\%, DT3: 17.2\%) less than the solution space (DT1: 37.1\%, DT2: 63.1\%, DT3: 36.9\%).

Overall, the results contradict Proposition A, that design thinking is not affected by geographic distance. We summarize this 
as:
\begin{displayquote}
\textbf{Observation 1:} \emph{Geographic distance affects design thinking by reducing problem space exploration, and alignment between problem- and solution space.}
\end{displayquote}

\subsection{RQ2: Challenges to Distributed Design}

While RQ1 is concerned with how \emph{design thinking} is affected by geographic distribution, RQ2 aims to answer why geographic distance affects \emph{design thinking}, i.e., what challenges designers perceive when working in a distributed fashion.
In the following, we present which challenges were perceived by the distributed teams in Case 2.

\begin{figure}[!t]
  \centering
  \includegraphics[width=0.4\columnwidth]{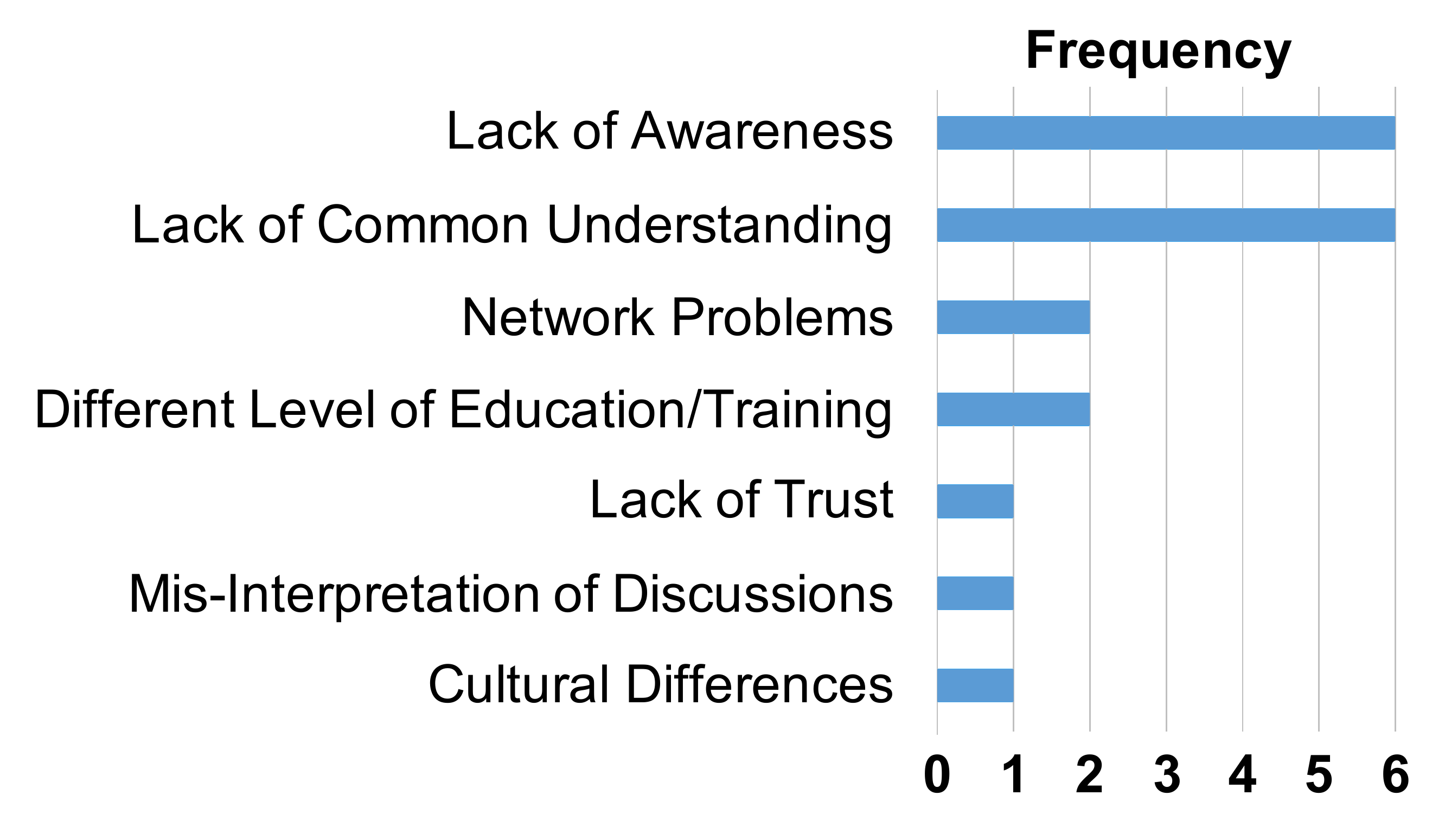}
  \caption{Perceived challenges to distributed design}
  \label{fig:challenges}
\end{figure}

Based on the post-study questionnaire (see results in Figure \ref{fig:challenges}) and our observations, we find that there are multiple challenges to designing in a distributed way.

All six participants in Case 2 found that it was challenging to be aware of the remote individual's reactions to interactions or the remote individual's focus on the joint work.
We also clearly observed this challenge during the design sessions. For instance,
\noindent
\says{Where do you draw?	}, \says{I don’t see what you draw now}, and  
\says{- I don’t see you drawing at the moment. - I drew out the left of the map}.
Configuring the environment with more video cameras so that facial and hand gestures of developers can be seen independently of their position could mitigate this issue.

The second challenge mentioned by all participants was a lack of common understanding.
 
Two participants mentioned network problems.
We report quotes from the participants regarding these problems:
\says{There is no connection anymore},
\says{For some reason the Skype connection is getting worse}, and 
\says{You are no longer connected, we see clients zero}.
Furthermore, technology contributes strongly to the awareness challenge, and could additionally contribute to misinterpretations of discussions (reported by one participant).

\subsection{RQ3: Creative Events}
We identify and report a number of \emph{creative events} during the design sessions of both the co-located and distributed teams.
We further analyze and discuss the dialogues or activities that lead to these creative events.

Tables~\ref{tab:CoCreaMom} and \ref{tab:DisCreaMom} report the observed creative events for the co-located and distributed teams respectively.
The second column in each table shows the frequency of creative events that relate purely to the problem space, to the solution space, or connect the two.
The third column then further breaks each type down into those events that were reached through divergent and those reached through convergent discussions.

\begin{table}
\parbox{.49\linewidth}{
\centering
\setlength{\extrarowheight}{0pt}
\addtolength{\extrarowheight}{\aboverulesep}
\addtolength{\extrarowheight}{\belowrulesep}
\setlength{\aboverulesep}{0pt}
\setlength{\belowrulesep}{0pt}
\caption{Creative events of the co-located teams}
\label{tab:CoCreaMom}
\begin{tabular}{llll} 
\toprule
\textbf{Team~~ } & \textbf{Event Type \textit{(Frequency)}~~ } & \textbf{Discussion \textit{(Frequency)}~~ } \\ 
\midrule \multirow{5}{*}{CT1} & Problem (0) &
\begin{tabular}[c]{@{}l@{}}\begin{tabular}{@{\labelitemi\hspace{\dimexpr\labelsep+0.5\tabcolsep}}l}Convergent (0)\\ Divergent (0)\end{tabular}\end{tabular}  \\
& {\cellcolor[rgb]{0.933,0.933,0.933}}Problem-Solution (7)  & {\cellcolor[rgb]{0.933,0.933,0.933}}\begin{tabular}{@{\labelitemi\hspace{\dimexpr\labelsep+0.5\tabcolsep}}>{\cellcolor[rgb]{0.933,0.933,0.933}}l}Convergent (7) \\ Divergent (0)\end{tabular}
\\ & Solution (18) &
 \begin{tabular}{@{\labelitemi\hspace{\dimexpr\labelsep+0.5\tabcolsep}}l}Convergent (13) \\ Divergent (5)\end{tabular} \\
 \midrule \multirow{5}{*}{CT2} & Problem (6) &
\begin{tabular}[c]{@{}l@{}}\begin{tabular}{@{\labelitemi\hspace{\dimexpr\labelsep+0.5\tabcolsep}}l}Convergent (4)\\ Divergent (2)\end{tabular}\end{tabular}  \\
& {\cellcolor[rgb]{0.933,0.933,0.933}}Problem-Solution (3)  & {\cellcolor[rgb]{0.933,0.933,0.933}}\begin{tabular}{@{\labelitemi\hspace{\dimexpr\labelsep+0.5\tabcolsep}}>{\cellcolor[rgb]{0.933,0.933,0.933}}l}Convergent (3) \\ Divergent (0)\end{tabular}
\\ & Solution (2) &
 \begin{tabular}{@{\labelitemi\hspace{\dimexpr\labelsep+0.5\tabcolsep}}l}Convergent (1) \\ Divergent (1)\end{tabular} \\
 \midrule \multirow{5}{*}{CT3} & Problem (0) &
\begin{tabular}[c]{@{}l@{}}\begin{tabular}{@{\labelitemi\hspace{\dimexpr\labelsep+0.5\tabcolsep}}l}Convergent (0)\\ Divergent (0)\end{tabular}\end{tabular}  \\
& {\cellcolor[rgb]{0.933,0.933,0.933}}Problem-Solution (2)  & {\cellcolor[rgb]{0.933,0.933,0.933}}\begin{tabular}{@{\labelitemi\hspace{\dimexpr\labelsep+0.5\tabcolsep}}>{\cellcolor[rgb]{0.933,0.933,0.933}}l}Convergent (1) \\ Divergent (1)\end{tabular}
\\ & Solution (8) &
 \begin{tabular}{@{\labelitemi\hspace{\dimexpr\labelsep+0.5\tabcolsep}}l}Convergent (5) \\ Divergent (3)\end{tabular} \\
\bottomrule
\end{tabular}
}
\hfill
\parbox{.49\linewidth}{

\centering
\setlength{\extrarowheight}{0pt}
\addtolength{\extrarowheight}{\aboverulesep}
\addtolength{\extrarowheight}{\belowrulesep}
\setlength{\aboverulesep}{0pt}
\setlength{\belowrulesep}{0pt}
\caption{Creative events of the distributed teams}
\label{tab:DisCreaMom}
\begin{tabular}{llll} 
\toprule
\textbf{Team~~ } & \textbf{Event Type \textit{(Frequency)}~~ } & \textbf{Discussion \textit{(Frequency)}~~ } \\ 
\midrule \multirow{5}{*}{DT1} & Problem (4) &
\begin{tabular}[c]{@{}l@{}}\begin{tabular}{@{\labelitemi\hspace{\dimexpr\labelsep+0.5\tabcolsep}}l}Convergent (2)\\ Divergent (2)\end{tabular}\end{tabular}  \\
& {\cellcolor[rgb]{0.933,0.933,0.933}}Problem-Solution (6)  & {\cellcolor[rgb]{0.933,0.933,0.933}}\begin{tabular}{@{\labelitemi\hspace{\dimexpr\labelsep+0.5\tabcolsep}}>{\cellcolor[rgb]{0.933,0.933,0.933}}l}Convergent (3) \\ Divergent (3)\end{tabular}
\\ & Solution (11) &
 \begin{tabular}{@{\labelitemi\hspace{\dimexpr\labelsep+0.5\tabcolsep}}l}Convergent (11) \\ Divergent (0)\end{tabular} \\
 \midrule \multirow{5}{*}{DT2} & Problem (4) &
\begin{tabular}[c]{@{}l@{}}\begin{tabular}{@{\labelitemi\hspace{\dimexpr\labelsep+0.5\tabcolsep}}l}Convergent (3)\\ Divergent (1)\end{tabular}\end{tabular}  \\
& {\cellcolor[rgb]{0.933,0.933,0.933}}Problem-Solution (4)  & {\cellcolor[rgb]{0.933,0.933,0.933}}\begin{tabular}{@{\labelitemi\hspace{\dimexpr\labelsep+0.5\tabcolsep}}>{\cellcolor[rgb]{0.933,0.933,0.933}}l}Convergent (4) \\ Divergent (0)\end{tabular}
\\ & Solution (5) &
 \begin{tabular}{@{\labelitemi\hspace{\dimexpr\labelsep+0.5\tabcolsep}}l}Convergent (5) \\ Divergent (0)\end{tabular} \\
 \midrule \multirow{5}{*}{DT3} & Problem (0) &
\begin{tabular}[c]{@{}l@{}}\begin{tabular}{@{\labelitemi\hspace{\dimexpr\labelsep+0.5\tabcolsep}}l}Convergent (0)\\ Divergent (0)\end{tabular}\end{tabular}  \\
& {\cellcolor[rgb]{0.933,0.933,0.933}}Problem-Solution (4)  & {\cellcolor[rgb]{0.933,0.933,0.933}}\begin{tabular}{@{\labelitemi\hspace{\dimexpr\labelsep+0.5\tabcolsep}}>{\cellcolor[rgb]{0.933,0.933,0.933}}l}Convergent (2) \\ Divergent (2)\end{tabular}
\\ & Solution (7) &
 \begin{tabular}{@{\labelitemi\hspace{\dimexpr\labelsep+0.5\tabcolsep}}l}Convergent (4) \\ Divergent (3)\end{tabular} \\
\bottomrule
\end{tabular}
}
\end{table}

\begin{figure*}[t!] \centering
\includegraphics[width=0.98\textwidth]{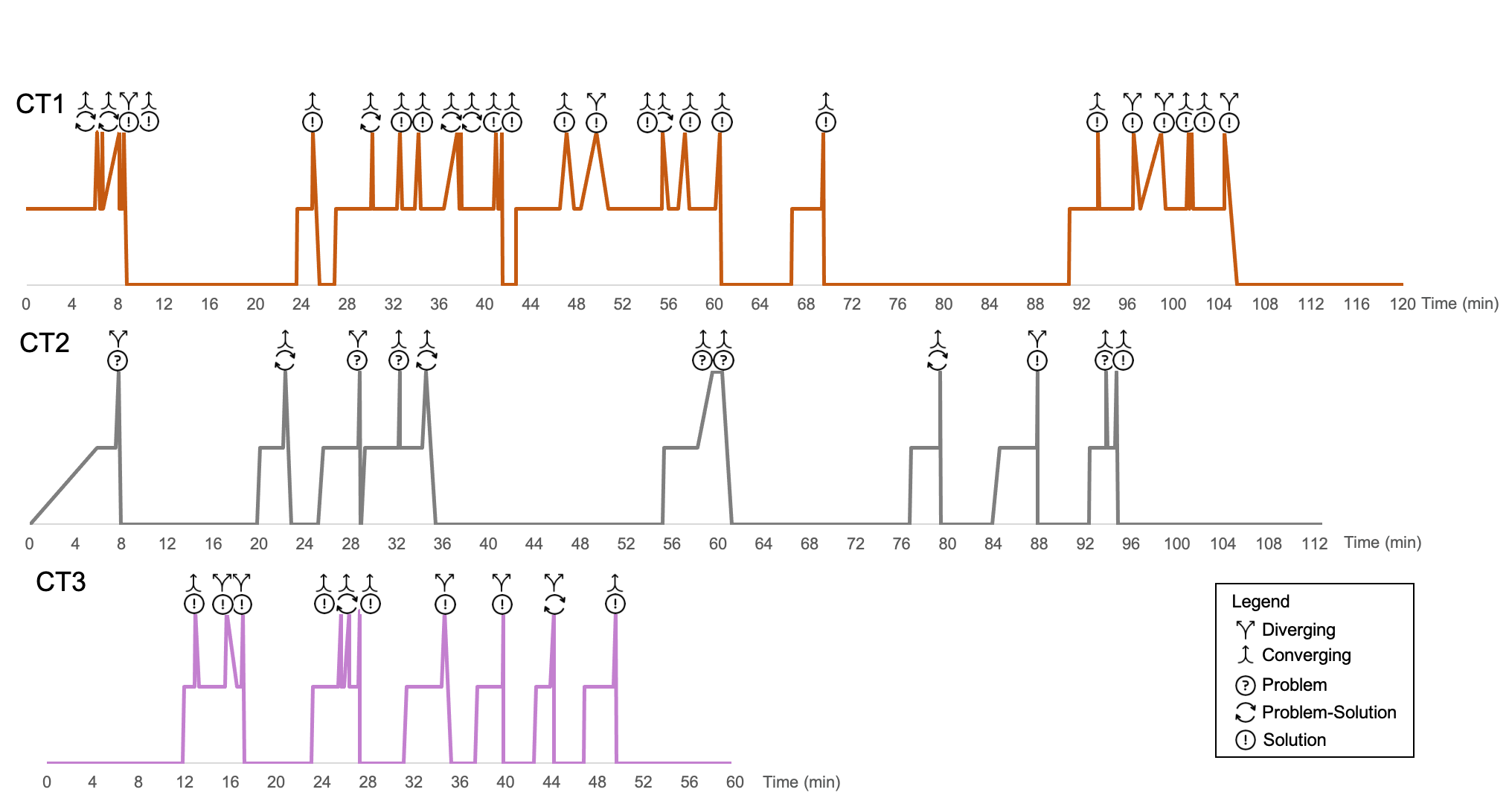}
\caption{Timeline of of the three co-located teams presenting their creative
events regarding problem space, solution space, and problem-solution bridges 
including discussions leading to these events.
Icons refer to the spatially closest peak}
	\label{fig:TimelineCT}
\end{figure*}

\subsubsection{Co-located Teams} 

The three co-located teams collaborated on a design task of up two $120$
minutes each.
Their creative events and the discussions leading to these are illustrated
in \autoref{fig:TimelineCT}.
Here, a each graph represents a conversation and structures it according to
creative events (graphical peaks), discussions leading to creative events
(elevated areas), and other discussion.
The discussions leading to creative events can be either convergent or
divergent~\cite{gabora2010revenge}.

The conversations feature between six and zero creative events relating to
the problem space, between two and seven creative events relating to
problem-solution bridges, and between two and $18$ creative events relating
to the solution space.
Out of these, between six and $20$ discussions leading to these events are
convergent and between three and five discussions are divergent.

Considering the different lengths of discussions relative to the number or
kinds of different events, this hints that quality and creativity are
independent of discussion length.
Moreover, the three conversations yield very different structures: the
occurrence of creative events begins after initial discussion of different
duration,the kinds of creative events and discussions are different and
their distribution is as well.

We report our observations on the discussions of the three teams below.

\subsubsection*{Co-located Team CT1} 

At the start of CT1's interaction, the team follows a pattern in which required concepts are enumerated in a quick order.
The first two creative events are problem-solution bridges that are directly linked to the requirements, e.g.,:
\says{Okay, so intersection, then we need car, we need the notion of time, this is a simulation of time, right.}.

Overall, CT1 does not seem to have any \emph{sudden} realizations or highly emotional moments. 
However, they engage in long dialogues about one single concept or issue, ultimately converging to a solution, a simplification, or an assumption: \says{Yeah, right away rules.} \says{Okay this is getting more complicated.} \says{It's coming together though, I think. It's just more rules.} \says{Yeah, yeah just need, well one probably also needs an admin right?}. 
In terms of our codes, this is visible in the absence of creative events that relate purely to the solution space, in a high proportion of solution events (18 out of 25), and in a majority of convergent discussions (20 out of 25).

On several occasions, the team takes real-world examples into account to reflect on their solution and to ask themselves whether the solution needs to be more complex, e.g.,:
\says{Do you want to keep it or do you want to do two queues?} \says{It depends if there's two lanes or not.} \says{Lanes...Yeah, the lane would be an addition of the road with some specific logic data.}.

\subsubsection*{Co-located Team CT2} 

CT2 discuss different aspects of the overall task at depth.
However, at several occasions, when encountering a complex or unclear issue, they make an assumption or decide that they keep it as a question to the customer later on: \says{This is something where I'd go back to the customer and try and figure out, how did they collect this data, you know like Professor E must have statistics about, you know San Francisco traffic}.
This enables the team to cover a large amount of topics during their session, but also to a conclusion of many topics at a shallow level, without deeper discussions.
Thus, the team hardly ever reaches any realizations.

Interestingly, CT2 has an event that comes closest to the definition by Dorst and Cross\cite{dorst01} towards the end of their design session.
After several assumptions and discussions referred to the customer, they finally seem to realize the true goal of the task: \says{Exactly, exactly. Because that's the purpose of Professor E's thing is to how do you correlate these settings with the result.}.
This realization allows the team to reason better about several of the assumptions and simplifications made earlier on.

Another interesting observation is that CT2 has the highest proportion of creative events relating to problem space and problem-solution bridges.
Indeed, only 2 creative events relate to the solution space, out of 11 overall.

\subsubsection*{Co-located Team CT3}

CT3 exhibits both phases of divergent and of convergent thinking.
In some cases, they switch frequently between topics.
For instance, in the beginning of their design session, the team quickly covers a lot of different topics, leading to a realization which part of the task is of major importance:
\says{So it seems like [..] designing intersections is kind of the key thing. I think so.} \says{Yeah, I agree.}
Later on, they re-visit the task in a structured way to see what parts are missing: \says{Okay cool, so now come back up, what big concepts haven't we captured yet?}.
Creative events are distributed evenly throughout their interaction, compared to most other teams that tend to have creative events predominately during the beginning and towards the end of their interactions.

\begin{figure*}[t!]
  \centering \includegraphics[width=0.98\textwidth]{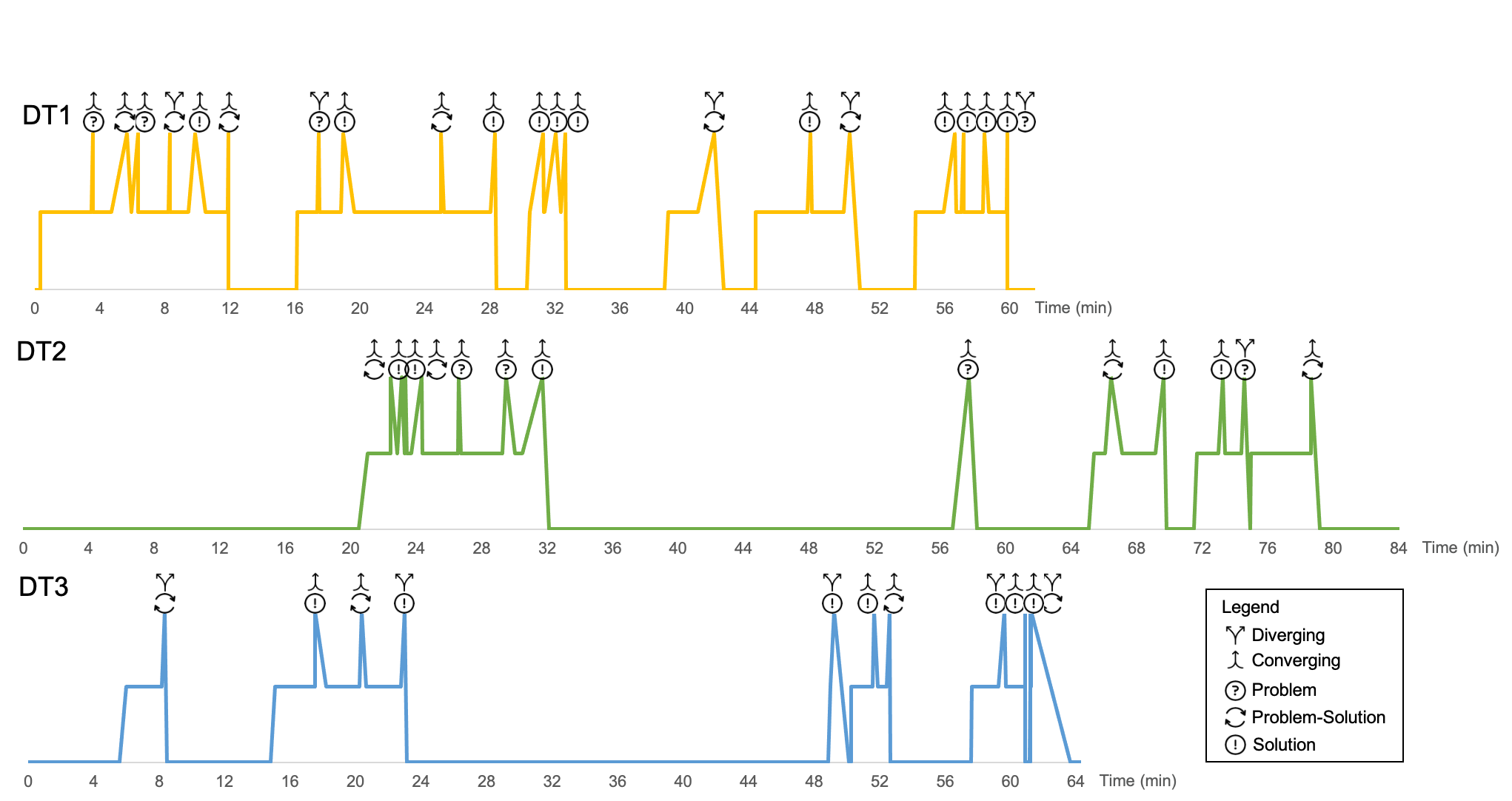}
  \caption{Timeline of of the three distributed teams presenting their
  creative events regarding problem space, solution space, and
  problem-solution bridges including discussions (converging or diverging)
  leading to these events. Not all creative events are represented
  graphically due to space limitations. Icons refer to the spatially closest
  peak}
  \label{fig:TimelineDT}
\end{figure*}
 
\subsubsection{Distributed Teams}
 
The three distributed teams collaborated on a design task of up two $60$
minutes each.
Their creative events and the discussions leading to these are illustrated
in \autoref{fig:TimelineDT}.
Again, a each graph represents a conversation and structures it according to
creative events (graphical peaks), discussions leading to creative events
(elevated areas), and other discussion.
The discussions, again, can be either convergent or
divergent~\cite{gabora2010revenge}.

The conversations yield between zero and four creative events regarding the
problem space, between four and six creative events regarding
problem-solution bridges, and between five and eleven creative events
regarding the solution space.
The discussion leading to these events are mostly convergent (between five
and $16$ discussions) and less often divergent (between one and five
discussions).

From this graphical overview, we can see that our results confirm Proposition A2, that design creativity is not affected by geographic distance. However, there are large differences between teams. We summarize this observation as:
\begin{displayquote}
\textbf{Observation 2:} \emph{Geographic distance does not clearly affect design creativity. Overall, there are large differences between teams in the frequency and location of creative events.}
\end{displayquote}

Below, we detail observations on the individual discussions.

\subsubsection*{Distributed Team DT1}

Developers in DT1 follow a strategy in which, initially, they iterate on what is required.
Therefore, the first 7 creative events are exclusively about what is required.
\says{But maybe first… so these are the things we need to do at real time.[..] } \says{Do you have something else? [..]} \says{I guess there should be some statistic [..]} \says{Is it mentioned in the task or do you just think it would be useful?}

After the initial exploration, the remainder of the discussion is mainly solution-focused.
Essentially, different requirements are addressed one by one, or "ticked off".
There is only very little problem focus during this period.
Indeed, DT1 has only a single creative event that is purely focused on the problem space: In the very end of their dialogue, they discuss whether any requirements are left.
\says{Okay then maybe let’s check again the requirements.}

The discussion is predominately convergent.
Divergent episodes are only found when the team tries to get an overview of the problem space (in the very end), or an overview of how concepts from the problem space are addressed in the solution.
 
\subsubsection*{Distributed Team DT2}

DT2 has structured discussions with little jumps between topics, mainly in a convergent fashion.
The only divergent creative event occurs towards the end of the session, when the team discusses any remaining requirements:
\says{Is there anything else what we are missing right now, despite the sensors?}
Sometimes, realizations arise out of the blue, but still seem to be connected to the ongoing (convergent) discussion.
For instance, at a point where the team discusses the topic of how to simulate intersections, one team member realizes that sensors have not been covered: \says{So do we, ah, ok what we are missing right now is the sensor}.

In contrast to, e.g., DT1, the team does not cover many parts of the problem domain, since they end up discussing the intersection and simulation at length. 
However, their discussions go to a much deeper level of the problem space, and fewer assumptions are made.
This is also evident in the large proportion of creative events that are connected to the problem space, or connected to bridges in the problem and solution spaces.
A technique the team regularly uses is to think through a real-life case and try to compare that to the task at hand:
\says{At least here in Sweden [..] the main road is always green if it doesn’t detect anything coming from the crossing roads. So the big road is always green, but when there are cars coming in from left or right it detects that and then it sets the main road to red so that they can join.}

\subsubsection*{Distributed Team DT3}

In DT3, several realizations in the beginning (after approximately 20 minutes) are essentially short-cut by making simplifying assumptions: \says{I would say so, but I would do this on some…We can rely on third party libraries. So there is a library that can randomly compute how the cars would move.}.
In particular, the assumption that simulation is covered by a third-party library avoids a discussion that took the majority of time in other teams.

Several minor realizations occur through enumeration of the concepts in the problem space.
Here, the team is trying to get an overview of what they have done so far.
This is a form of divergent thinking, jumping between concepts in a seemingly defocused fashion:
\says{Should we model any more parts? I mean there needs to be some kind of simulation engine [..] And we need probably something like some statistic engine [..] And UI of course.}.

An additional style exhibited by DT3 is the presence of creative events that map problem space to solution space in a convergent manner.
That is, the team jumps between requirements (problem space) and UI or implementation details (solution space) to understand which concepts from the problem space already exist in the solution space:
\says{Okay. What did you think of, of. What did you envision [..]} \says{it is [sic] enough to just have some a kind of a grid of with varying lengths between the intersections. What could that be? [..]} \says{We still need the item pane at the bottom for having traffic light or these sensor thingies or whatever it was in the requirements.}.
\label{sec:discussion}

In this section we discuss the results in relation to existing work. We discuss results of RQ1 in Section~\ref{sect:discRQ1}; RQ2 in Section~\ref{sect:discRQ2}; and RQ3 in Section~\ref{sec:discRQ3}.

\subsection{RQ1 Geographic Distance Affects Design Thinking}
\label{sect:discRQ1}

Our results provide a potential explanation as to how design communication is hampered in geographically-distributed teams, namely by reducing problem space exploration in favor of a more solution-oriented communication.
Clark et al. \cite{clark1991grounding} state that in a collaborative setting, individuals keep on discussing and sharing knowledge until they reach a mutual understanding of the discussed argument.
In problem space exploration, developers extensively exchange and complement their knowledge of the domain in order to reach a shared understanding of the problem space \cite{cross2004expertise}. 
Achieving a shared understanding of the problem space might therefore be hampered in geographically-distributed teams.
This is an interesting addition to the more general observation that geographic distance can hamper communication and coordination~\cite{bjarnason2016theory,herbsleb2007global,damian2007collaboration}.

In summary, we observe that geographic distribution reduces problem space exploration in collaborative software design tasks compared to the co-located setting, thus answering RQ1.
We believe that this finding explains in more detail how communication is affected by geographic distance.

\subsection{RQ2 Technology and Social Factors, the Common Challenges to Designing in a Distributed Way}
\label{sect:discRQ2}

We have observed that gestures or facial expressions were hard or impossible to recognize.
This was the case despite the provided video-conferencing facilities, e.g., when they were moving in front of the interactive whiteboard.
This is a technological challenge known from related work on distributed work~\cite{dourish1992awareness}.

Lack of common knowledge, challenge referring to individuals not knowing whether they have the same level of knowledge or
understanding of the subject matter as their remote counterpart \cite{herbsleb2007global} was encountered.
In the Gap model \cite{bjarnason2016theory}, this challenge relates to cognitive distance.
This issue could be reduced by collaborating with developers from the same department or school of thought.

The results of the post-study questionnaire confirm Proposition B (tool support and social factors are the most frequent challenges in GSE).
However, socio-cultural factors might be under-represented, since participants are likely more aware of technical limitations, e.g., due to cognitive bias.

It is interesting to note that only one participant noted cultural differences as a perceived challenge.
In Case 2, German developers were partnered with Swedish developers.
As the German and Swedish cultures do have large differences on the Hofstede culture dimensions \cite{hofstede2001culture}, especially in masculinity and uncertainty avoidance, we would have expected a more noticeable influence of culture.
However, as noted above, participants might not have perceived this challenge strongly.

In addition to socio-technical factors, the Gap model \cite{bjarnason2016theory} includes artifact-related distances, e.g., semantic distances between multiple specifications.
In our study, these distances were excluded by design and therefore do not show up in the perceived challenges.
However, even if they were included, we believe that they would have a similar influence in both co-located and distributed settings.

In summary, we can answer RQ2 stating that both technical and social challenges are commonly encountered in collaborative, distributed software design.

While we cannot directly infer a causal connection between our results for RQ1 and RQ2, it is likely that the observed challenges for RQ2 play a major role in how \emph{design thinking} is affected in Case 2.
To be able to efficiently explore problem and solution space, and to align them iteratively, it is essential to be aware of the current situation.
However, the top two challenges, a lack of awareness of the remote counterpart and a lack of common understanding, indicate that this is not sufficiently possible.
This can lead to less overall communication, or lead to more explicit communication, e.g., by not exploring the problem space sufficiently (as noted in \textbf{Observation 1}).
The observed reduction in problem space exploration could be explained in two ways.
First, individuals typically rely on previous knowledge for problem solving instead of a thorough analysis \cite{christiaans2010accessing}.
This effect might be amplified by socio-technical barriers.
Second, due to different barriers introduced by geographic distance, developers in Case 2 might do more tacit, internal \emph{design thinking} and only present ideas after a number of internal iterations through the \emph{design thinking} loop. This could lead to a lower overall quality of the resulting product, since not all ideas or solution candidates are discussed by all developers involved in the collaboration.
For example, the distributed developers did not have a direct face-to-face communication and, therefore, missed eventual facial expressions and hand gestures, which often give a hint whether or not an argument is mutually understood. 

Overall, we formulate the causal relation between our findings for RQ1 and RQ2 as a hypothesis for future work:
\begin{displayquote}
\textbf{Research Hypothesis 1 (RH 1):} \emph{Lack of awareness of the remote counterpart and lack of common understanding reduce the amount of problem space exploration in distributed \emph{design thinking}.}
\end{displayquote}

\subsection{RQ3 Geographic Distance Does not Clearly Affect Design Creativity}
\label{sec:discRQ3}

According to Dorst and Cross~\cite{dorst01}, creative events are "highly emotional steps", where subjects feel they have understood or uncovered "the underlying problem" behind or "the difficulty" with a task.
Dorst and Cross~\cite{dorst01} observed such events in all their study participants.
In contrast, we did not observe events that clearly followed this pattern in any of the teams.
Therefore, we relaxed the coding guidelines towards a more general notion of insights or realizations regarding the problem or the solution, as discussed in Section~\ref{sec:approach}.

Even with the relaxed coding guidelines, we notice an overall low amount of divergent discussions leading to creative events, compared to the amount of convergent discussions that result in realizations.
As noted by Goldschmidt~\cite{goldschmidt16} and Gabora~\cite{gabora2010revenge}, a combination of divergent and convergent discussions is believed to increase design creativity.
Hence, this indicates that our teams exhibit relatively little design creativity, and that encouraging more divergent discussions could increase their creativity.
We formulate this potential improvement as another hypothesis for future work:
\begin{displayquote}
\textbf{RH 2:} \emph{Encouraging more divergent discussions in software design tasks increases design creativity.}
\end{displayquote}

There are several potential arguments why we did not observe creative events as described by Dorst and Cross, and why divergent discussions were comparably few.
First, our study design might have been too vague, broad, or permissive so that true insights were restricted.
For instance, given the broad scope of the task, participants might have generally preferred a shallow exploration of the problem, covering a large proportion of the requirements in favor of a deeper discussion.
Secondly, Dorst and Cross~\cite{dorst01} study industrial design, as opposed to this SE task.
That is, the domain and the subject of software design could inhibit creative events as described by Dorst and Cross.
One argument for this option is that software design is on a high level of abstraction, trying to avoid detailed implementation decisions.
In the given task, this level of abstraction could make it possible that participants stop the discussions before reaching the complexity necessary to truly understand the underlying difficulties.
Similarly, the multiple different viewpoints from which a software design task can be addressed in contrast to a purely physical device could make the software design process different from industrial design.
Thirdly, in our setup, pairs of participants are discussing a design task, without the possibility to ask intermediate questions.
In contrast, Dorst and Cross~\cite{dorst01} study a setup in which individual participants solve a design task in which they are able to interact with an \emph{expert}.
In several of our teams, e.g., DT1 and DT3, we noticed that participants made assumptions or decided to defer the issue at points in the discussion where the complexity of the problem became apparent.
For instance, CT2 had several statements similar to the following excerpt, after which they would make an assumption or continue with another topic:
\says{Yeah, I think we're going to have to rely on professor E for creating the details about the theory of how that works.}.
Having the option to clarify would have either solved the issue, or given the team enough information to continue discussions, potentially leading to "true" creative events.
Finally, as indicated by Mohanani et al.~\cite{mohanani19}, framing desiredata as requirements can inhibit design creativity, something that indeed could have been the case in our task.

Regarding the connection between design creativity and design thinking, we observe that the teams that quickly jump between topics and make many assumptions seem to be the teams that exhibit most problem space exploration.
In the co-located case, CT2 has with 43.8\% by far the highest proportion of problem-space exploration, followed by CT1 with only 29.4\%. 
Investigating the creative events, we notice that this team exhibits a clear pattern of making assumptions whenever they come across a new issue or complexity that they cannot solve immediately:
\says{So this can get fairly complex very fast because you have all these intersections. [..] 
Lets assume this, lets assume it's inside a city so...[..] We'll talk to Professor E about it.}.
This allows the team to quickly cover a large amount of concepts from the problem space, potentially leading to the high proportion of problem-space exploration.
In the distributed case, DT1 has 32.1\% problem-space exploration, followed by DT3 with 17.2\%.
As for CT2, DT1 makes a large number of assumptions, without going in depth with the respective issue.
In the analysis of creative events, this is reflected by a majority of events that relate to the solution space.
All events that purely relate to the problem space are events in which a participant reads a requirement and realizes that a concept/a requirement is missing in the solution, e.g.,:
\says{Here it says: Students must … with or without the option to have sensors. So it’s stated that it is related to the intersection, not to the road.}
There is no evident correlation between the number of creative events and the proportions of problem and solution space exploration.
Similarly, we do not see any correlation between the number of convergent/divergent discussions leading to a creative event.
This indicates that creative events cover a dimension orthogonal to the design thinking phenomenon.

A final observation worth discussing is the notion of strategies how to reach creative events.
Dorst and Cross~\cite{dorst01} state that designers seem to have a strategy to reach creative events: "They search through the information by asking a quasi-standard set of questions, such as: 'capability of the company', 'available investment', etc. Apparently, they have a set of expectations about the answers to these questions. [..] In doing this, they check the information related to the assignment to build up a general image, and to look for surprises."~\cite{dorst01}.
In our teams, we observe a number of similar strategies.
First, enumeration of problem/solution space concepts is used by most teams at the beginning and the end of their sessions to find complex or missing concepts.
While the teams start the task with an expectation of what is needed, this enumeration helps them to find any surprising concepts that they did not initially consider, or that are complex to translate to the solution space.
At the end of their sessions, the enumeration of problem space concepts is used as a form of reconciliation, to identify whether any obvious concepts or requirements were missed or not addressed in the solution space.
Finally, we notice several teams taking into account real-life situations when they are stuck in the solution space.
That is, if there is an issue they cannot directly solve, they try to understand how this works in reality, in the hope that this will help them solving their problem.

 \section{Threats to Validity}
\label{sec:threats}
In the following, we will describe potential threats to validity and how we addressed them in our study.
We follow the classification into different threats according to Maxwell~\cite{maxwell92}, as they fit a broad spectrum of different scientific world views~\cite{petersen13}.
For an account how this classification compares to more common ways of discussion validity in SE, such as internal, external, conclusion, and construct validity, please refer to Petersen and Gencel~\cite{petersen13}.

\subsection{Descriptive Validity}
Descriptive validity in qualitative research is the "factual accuracy" of the reported data~\cite{maxwell92}.
It is not concerned with any interpretation of the data, only whether it is accurately represented in the account.

In our study, we used clean verbatim transcripts of the recorded video sessions.
As such, accounts of data are not relying on memory.
However, there is a minor threat to descriptive validity since we used transcripts for data analysis instead of the video recordings.
This could have affected our analysis, since the tone of the voice, gestures, or facial expressions might have contained important information on emotional state or emphasis.

\subsection{Interpretive Validity}
Interpretive validity concerns whether "conclusions/inferences drawn [are] reasonable given the data representing an objective/subjective truth"~\cite{petersen13}.
That is, in contrast to descriptive validity, the interpretation of data is a central concern.

To reduce threats to interpretive validity, we employed a number of measures.
First, we relied on two established theoretical concepts: the design thinking concept \cite{cross2011design} and that of convergent/divergent thinking \cite{goldschmidt16}.
For coding the data, we used written guidelines such as the coding guidelines presented in Appendix~\ref{app_creativeEvents}.
Multiple researchers then applied these guidelines to the same data, and discussed differences.
In cases were it proved difficult to exactly apply the guidelines, we relaxed them.
For example, this was the case for creative event coding, as emotions and emphasis were not clearly identifiable in the written transcripts.

Furthermore, we described the case study design and process in detail in Section \ref{sec:approach}, to allow readers to assess the validity in more depth.
Finally, we use the published task description by \cite{petre2013software}, whose material can be obtained at \url{https://www.ics.uci.edu/design-workshop/videos.html}.
This enables researchers to replicate our study.
To the extent this is feasible in qualitative case studies, this should enable a reproduction of our results under comparable contexts.

\subsection{Theoretical Validity}
Theoretical validity refers to the validity of the theoretical concepts employed in a study and the relationships between them \cite{maxwell92}.

\emph{Design thinking} is a concrete phenomenon, a natural and ubiquitous human activity during problem-solving processes \cite{cross2001designerly}. 
We assessed \textbf{explicit} \emph{design thinking}, i.e., \emph{design thinking} expressed verbally.
However, \emph{design thinking} is a cognitive activity and can happen also implicitly inside the mind of the thinking developer (tacit \emph{design thinking}). 
This is a threat to theoretical validity, as the \emph{design thinking} processes captured in our study might not represent the overall \emph{design thinking} activity.

We collected data on challenges to distributed \emph{design thinking} by asking the distributed developers to report their perceived challenges during the design sessions.
These challenges are subjective and may vary from one developer to another.
Furthermore, developers might have forgotten to report some of the challenges that they experienced during the design session.
To mitigate this issue, we additionally tried to triangulate the reported challenges with those observed during the distributed design sessions.

The heterogeneity and selection of subjects is a further threat to theoretical validity.
While all teams consisted of members with professional experience, this experience differed.
Specifically, the co-located teams had on average 19 years of professional experience, the distributed teams between 3 and 7 years.
The distributed teams were recruited among personal contacts of the authors, while we did not have control over the recruitment in Case 1.
This might have affected how the teams interacted, and how they solved the task.
While a more homogeneous sample would have been beneficial, recruiting professionals for SE studies is a challenge.

As a last threat to theoretical validity, the theoretical constructs we used in our study might not have been defined well enough to ensure similar understanding among all researchers.
To reduce this threat, we used written coding guidelines as discussed for interpretive validity.
The coding schema used for the analysis of design thinking was furthermore based on an existing schema~\cite{weinreich2015expert}.
Furthermore, we regularly discussed intermediate coding results.

\subsection{Internal Generalizability}
Internal generalizability is the extent to which generalization "within the community, group, or institution studied to persons, events, and settings that were not directly observed or interviewed" \cite{maxwell92} is possible.

The level of expertise and experience in software architecture design might influence the \emph{design thinking}.  
Novice and expert designers are observed to have different strategies to solve ill-defined problems \cite{cross2004expertise,ahmed2003understanding}.
Experts are able to store and retrieve information in larger cognitive chunks than novices. 
Moreover, experts concentrate on recognizing underlying principles rather than focusing on the surface feature of problems. 
The results of this study expose and explain the \emph{design thinking} of expert developers. 
To understand the behavior of novice developers, we call for replication. 

Different developers might perceive different challenges to their GSE experience. 
The reported GSE challenges by the participants of Case 2 are in line with the challenges reported in GSE literature and practice.
However, we cannot guarantee that other developers would report the same challenges.

The participants in Case 2 used an interactive whiteboard (i.e., OctoUML) for their distributed collaboration.
This could have affected our comparison of the \emph{design thinking}. 
Moreover, using new tools often causes a learning effort.
To mitigate these two issues, we customized OctoUML to resemble a regular whiteboard as much as possible.
Furthermore, we deployed OctoUML on an interactive whiteboard and made use of Skype for live-communication. 
This allowed the developers to collaboratively communicate and concurrently sketch and at the same time.
In addition, all distributed developers had a short hands-on experience (i.e., $2$-$3$ minutes) to test the environment by collaboratively sketching on the shared canvas of OctoUML.
As discussed in Section \ref{sec:approach}, we measured the usability of OctoUML through a standard SUS questionnaire \cite{brooke2013sus}.
The average SUS score of $74.17 \pm 5.63$ (good usability according to \cite{sauro2011practical}) indicates that OctoUML has a reasonable usability and does not perform significantly worse than a state-of-the-art tool for distributed collaboration.
Hence, the use of OctoUML should not present a confounding factor in our study.
However, we also observed several discussion episodes in Case 2 that were interrupted by tool and connection problems.
This might have affected the design creativity of the teams, as they were interrupted in their thinking.
 
The design sessions lasted about $90$ minutes, i.e., some teams finished the task early even though they had more time. 
Thus, the participants might have suffered from fatigue that could have led to less \emph{design thinking} or design creativity.
We assume that fatigue would have a similar effect in both cases, and therefore not affect our comparison.

\subsection{External Generalizability}
External generalizability is the extent to which generalization "to other communities, groups, or institutions" \cite{maxwell92} is permitted by the study.

By design, case studies have a very limited external generalizability stemming from the fact that a topic is studied within its context, and that there is little to no control. 
Therefore, we cannot claim that our findings are generalizable, i.e., different projects in different domains might have different results.
However, we described the case context as detailed as possible in order to allow practitioners to decide whether or not the findings might apply in their own case context.

We asked participants to perform a specific software architecture design task.
In industrial settings, design tasks can vary substantially, e.g., in terms of size, terminology, language, and level of detail. In particular, our assignment is a green-field case - i,e, involving both the analysis of the problem domain and the synthesis of a solution from scratch.
This also limits the external generalizability of the findings. 
Moreover, we underline that our study involved expert software architecture designers. 
Different results might be obtained if novice designers were involved in this study.
Notwithstanding the aforementioned threats, many of our findings are consistent with the findings of other research on global software development.
 \section{Conclusion}
\label{sec:conclusion}
In this paper we reported a multiple-case study exploring the effect of geographic distance on design thinking (RQ1), the challenges perceived by geographically-distributed teams (RQ2), and the effect of distance on design creativity.
We observed that distributed developers did less \emph{design thinking} than the co-located developers, specifically in problem space exploration and in the alignment between the problem and solution space (\textbf{Observation 1}).
Participants in the distributed case perceived a lack of  awareness of the remote counterpart and a lack of common understanding as the main challenges in distributed design.
We hypothesize (\textbf{RH 1}) that these perceived challenges might be causing the observed reduction in \emph{design thinking}.
However, we did not observe obvious differences in design creativity between co-located and distributed teams (\textbf{Observation 2}), apart from connection issues that caused interruptions in the distributed teams.
Finally, we observe almost no creative events that strictly follow the description by Dorst and Cross~\cite{dorst01}, and only few divergent discussions leading to creative events.
We hypothesize (\textbf{RH 2}) that guiding teams to increase the amount of divergent discussions could increase design creativity, and therefore the amount of creative events.

\subsection{Implications for Research}
Our study extends the body of knowledge in GSE, since it offers an explanation how geographic distance affects communication, namely by reducing problem space exploration.
While it is known that GSE might hamper communication, our findings concretize this knowledge.
This is especially valuable since our analysis of \emph{design thinking} does not rely on subjective perceptions,
and can therefore complement data based on participant perceptions.
Similar analyses could be used in future work to measure the effect of potential solutions for designing in distributed settings.
To this end, we developed a general coding schema for analyzing design discourses, based on existing work by Weinreich et al.~\cite{weinreich2015expert}.

Our findings for RQ2 mainly confirm existing work in GSE.
Whether or not there is a causal connection between our findings for RQ1 and RQ2 is as of now hypothetical (\textbf{RH 1}).
Specifically, our findings rely mainly on the perceptions of study participants.
Since social challenges in particular might be implicit and therefore not visible in a self-evaluation, we see the need for studying this connection in more detail, i.e., how socio-technical barriers, such as culture or beliefs, affect \emph{design thinking} in distributed teams.
Increasing the understanding of these factors would contribute to a more efficient and effective remote collaboration, and thus result in products of higher quality.

Our findings for RQ3 offer mainly methodological insights for SE research.
In contrast to \emph{design thinking}, we do not observe any clear differences in design creativity between co-located and distributed teams, and no apparent correlation between the creative events and the observed \emph{design thinking}.
Factors that could play a role here are differences between SE and industrial design, which is the context in which they were described by Dorst and Cross~\cite{dorst01}, the abstract and multi-view nature of software design in contrast to purely physical devices, or methodological shortcomings of our study.
However, it could indeed be the case that problem-space exploration is reduced in distributed teams, but design creativity is not.
Hence, the distributed teams could simply be more effective in their problem-space exploration.
To our knowledge, there exists not previous work in SE that analyses design creativity by identifying creative events, and only little work that focuses on design creativity, e.g., Mohanani et al.~\cite{mohanani19}.
Therefore, similar studies are needed to evaluate the usefulness of this analysis approach.

Another interesting observation for RQ3 is the relative lack of divergent discussions that lead to creative events in our teams.
We hypothesized in previous work, which is extended by this paper, that future work could build on the reminder cards proposed by Tang et al.~\cite{tang2018improving}, to suggest a set of reminder cards that foster problem space exploration.
Our findings suggest that encouraging more divergent discussions should be a another focus of such a new approach, as it has the potential to increase overall design creativity in both co-located and distributed teams (\textbf{RH 2}).
The strategies for reaching creative events mentioned by Dorst and Cross~\cite{dorst01} could be another addition to such an approach.

\subsection{Implications to Practice}
While practitioners might be aware of negative effects of geographic distribution, our findings can help them understand how this manifests in practice.
This knowledge can in turn help them to decide how to engage in GSE.
A potential decision might be to limit tasks that require substantial problem space exploration to co-located teams, e.g., when features or products with large uncertainty are designed.
Similarly, in distributed design sessions, practitioners might decide to counteract the technological challenges observed in this study, e.g., by adding video conferencing that shows the faces and gestures of the involved designers.
These decisions could positively affect development achievement and, therefore, product quality and customer satisfaction.

In addition to technical factors, cultural and other social barriers might increase due to distribution.
Our findings confirm existing works with respect to the importance of these non-technical factors and can reinforce practitioners in initiatives that aim to mitigate these barriers.

\subsection{Future Work}
In addition to the several directions for future work outlined in this paper, we plan to extend our study to further SE activities.
In particular, we plan to analyze whether problem space exploration is reduced in a similar fashion in software modeling and requirements engineering activities.
In addition, we want to further investigate the socio-cultural dimension in this context, e.g., by controlling for different national cultures or knowledge gaps between different geographic sites.
Finally, our findings for RQ3 questions the extent to which the \emph{design thinking} and design creativity analyses are useful in SE.
We believe further studies in this direction need to be conducted, in order to clearly evaluate the potential of these theories in SE.

\section*{Acknowledgments}
We would like to acknowledge the subjects participating in our study for their valuable time. Furthermore, we thank André van der Hoek for sharing the dataset for the co-located teams with us.

\subsection*{Author contributions}
All the authors contributed in writing and reviewing this manuscript.  
The design of the ``distribution'' case study was performed by Rodi Jolak and Michel R.V. Chaudron.
The ``distribution'' case study was conducted by Rodi Jolak and Andreas Wortmann.
The two case studies were analyzed and discussed by Rodi Jolak, Andreas Wortmaan, Grischa Liebel, and Eric Umuhoza.

\subsection*{Financial disclosure}
None reported.

\subsection*{Conflict of interest}
The authors declare no potential conflict of interests.

\section*{Supporting information}
None reported.

\appendix
\section{Coding of Creative Events}
\label{app_creativeEvents}
Dorst and Cross~\cite{dorst01} discuss that bridges between the problem and solution space are built by identifying "key concepts".
The authors write that "creative design involves a period of exploration in which problem and solution spaces are evolving and are unstable until (temporarily) fixed by an emergent bridge which identifies a problem-solution pairing. A creative event occurs as the moment of insight at which a problem-solution pair is framed"~\cite{dorst01}.

We will aim to identify these key events (called \emph{Creative Events} in the following) by manually going through the textual transcripts.
The identification is a subjective process that cannot be complete formalized.
For instance, keywords have only little value as they are depending on communication style and language.
The following points are indicators that a Creative Event occurs:
\begin{enumerate}
    \item Subjects understand a connection that they were unawares beforehand. This simplifies an issue they were dealing with so far, resulting in a solution to a part of the assignment/task.
    \item In contrast to (1), subjects identify an issue they were unawares before, leading to a complication of the overall task.
    \item Subjects find an appealingly simple solution to an issue. This greatly simplifies the issue or the overall task.
    \item In terms of keywords, the subjects show sudden realization (e.g., "Aha", "Now I understand it", "Wait a moment, we haven't thought about..."). However, it is important to keep in mind that this is context dependent and can be hard to identify in the written transcriptions (e.g., a matter of fact "Now I understand what you are saying" vs. an emotional "Now I finally understood the assignment"). Dorst and Cross \cite{dorst01} state that Creative Events are generally "highly emotional steps".
    \item Regularly, the subjects feel that they have understood or uncovered "the underlying problem" behind or "the difficulty" with the task.
\end{enumerate}

After all Creative Events have been identified in all transcripts, we will look at the area around (+/- 5 minutes) each Creative Event and try to characterize it.
That is, we try to answer "What kind of discussion lead to the Creative Event?" and "What implications does the Creative Event have?".
To do so, we will use descriptive coding, i.e., we use codes similar to "hash tags" that describe what the subjects are discussing.
We will not use any pre-defined (a priori) codes, but rather use an open coding approach.
Similarly, we do not define the stanza (the unit which is coded), but leave it up to the coder to decide which parts to assign a code to.
This can also vary (e.g., one remarkable sentence gets a single code, while in another case only an entire paragraph is coded, and in yet another case, an entire paragraph does not get any code since it does not appear to be relevant for answering the questions). \end{document}